# Method for rapid estimation of the energy-time covariance matrix of single electrons

Wanki Park[1*†], Chanuk Yang[2,3†], Young-Seok Ghee[1], Hyung Kook Choi[3], Bum-Kyu Kim[1] and Myung-Ho Bae[1,4*]

[1]Korea Research Institute of Standards and Science, Daejeon 34113, Republic of Korea

[2]Jeonbuk National University G-LAMP Project Group, Jeonbuk National University, Jeonju 54896, Republic of Korea

[3]Department of Physics, Jeonbuk National University, Jeonju 54896, Republic of Korea

[4]KAIST Graduate School of Quantum Science and Technology, Korea Advanced Institute of Science and Technology, Daejeon 34141, Republic of Korea

[†]W.P. and C.Y. contributed equally to this work.

*e-mail: wanki931@gmail.com, mhbae@kriss.re.kr

**Abstract**

The ability to emit and control single electrons in a dynamical manner enables their use in electron quantum optics and sensing. To characterize the electron states emitted with energy far above the Fermi energy, a dynamic barrier has been used. In this work, we extract the energy-time covariance matrix of single electrons by analyzing the energy variance obtained from the transconductance through the dynamic barrier. This method enables efficient and precise characterization of electron states, especially when a sinusoidal waveform is used. An effective phase-space area and the elliptical distribution are constructed from the covariance matrix and qualitatively compared with the distribution reconstructed from a tomographic method. The area constrained by the uncertainty relation serves as a measure of proximity to the quantum limit. Our results demonstrate an efficient method for characterizing electron states, paving the way for their application in quantum technologies.

Single electrons in a two-dimensional electron gas (2DEG) system have been emitted via several methods such as time-varying modulations of a mesoscopic capacitor [1], electrical contact [2,3], interacting quantum dot (quantum dot pump) [4-7], and surface acoustic wave [8,9]. The emitted single electrons are used in applications such as sensing [10,11], electron quantum optics [3,12-14] and quantum information processing [15,16]. A crucial step in applying such system is measuring the quantum states of the emitted electrons, enabling verification of the prescribed states and evaluation of their suitability for the aforementioned applications. Various techniques have been employed to characterize the electron states by manipulating them with a beam splitter formed by quantum point contacts, with the specific method depending on the emission energy of the electrons.

For low-energy electrons, such as those excited from electrical contacts or mesoscopic capacitors, their quantum states are typically characterized using Hong-Ou-Mandel-type collisions at the beam splitter with a reference Fermi sea [17-19]. Energy-resolved measurements are achieved by tuning the Fermi energy, while time-resolved measurements are facilitated by applying a sinusoidal voltage to modulate the Fermi level, enabling reconstruction of the electron density matrix [17-19].

For high-energy electrons, significantly above the Fermi energy, such as those emitted by a quantum dot pump, their states are characterized using a potential barrier (detector) that acts as an energy filter, allowing only electrons with energies exceeding the barrier height to transmit. Energy-resolved measurements are conducted by measuring the transmitted current while varying the barrier height [20-22]. Time-resolved measurements are achieved by modulating the barrier height over time, as detailed later.

Fletcher et al. [21] applied a sinusoidal voltage to the potential barrier, enabling temporal access and allowing the estimation of single-electron arrival times and phase-space distributions. This method was further developed into a continuous-variable tomographic approach [23,24], where projection images of the phase-space distribution are obtained by applying a linear voltage to the barrier and tuning its slope [25]. These projection images allow for the reconstruction of the Wigner distribution, which offers a phase-space representation of the electron state equivalent to the density matrix. It provides the purity of an electron state, reflecting how mixed the quantum state is. The distribution is also estimated under the assumption of an elliptical shape [26].

In this Letter, building on the prior works, we introduce an efficient yet precise method to estimate the quantum state of high-energy electrons. We develop a formula to extract the covariance matrix, with energy and time variances on the diagonal and their covariance on the off-diagonal. This is obtained by analyzing variations in measured energy variance of the transconductance as a function of the barrier ramping rate. Measurements are performed by applying both linear or sinusoidal gate voltage waveforms to the barrier. Our results show good agreement between the two cases, both widely utilized in previous studies. The sinusoidal waveform is efficient in our framework, enabling shorter measurement times without requiring high-frequency waveform synthesis, in contrast to the continuous-variable tomographic method. We use the covariance matrix to construct an elliptical distribution and define an effective phase-space area, which qualitatively matches the distribution from tomography. The area is found to be $\sim 20\hbar$, much larger than the quantum limit of $\hbar/2$. We suggest that this area can serve as a practical measure of the mixedness of the quantum state.

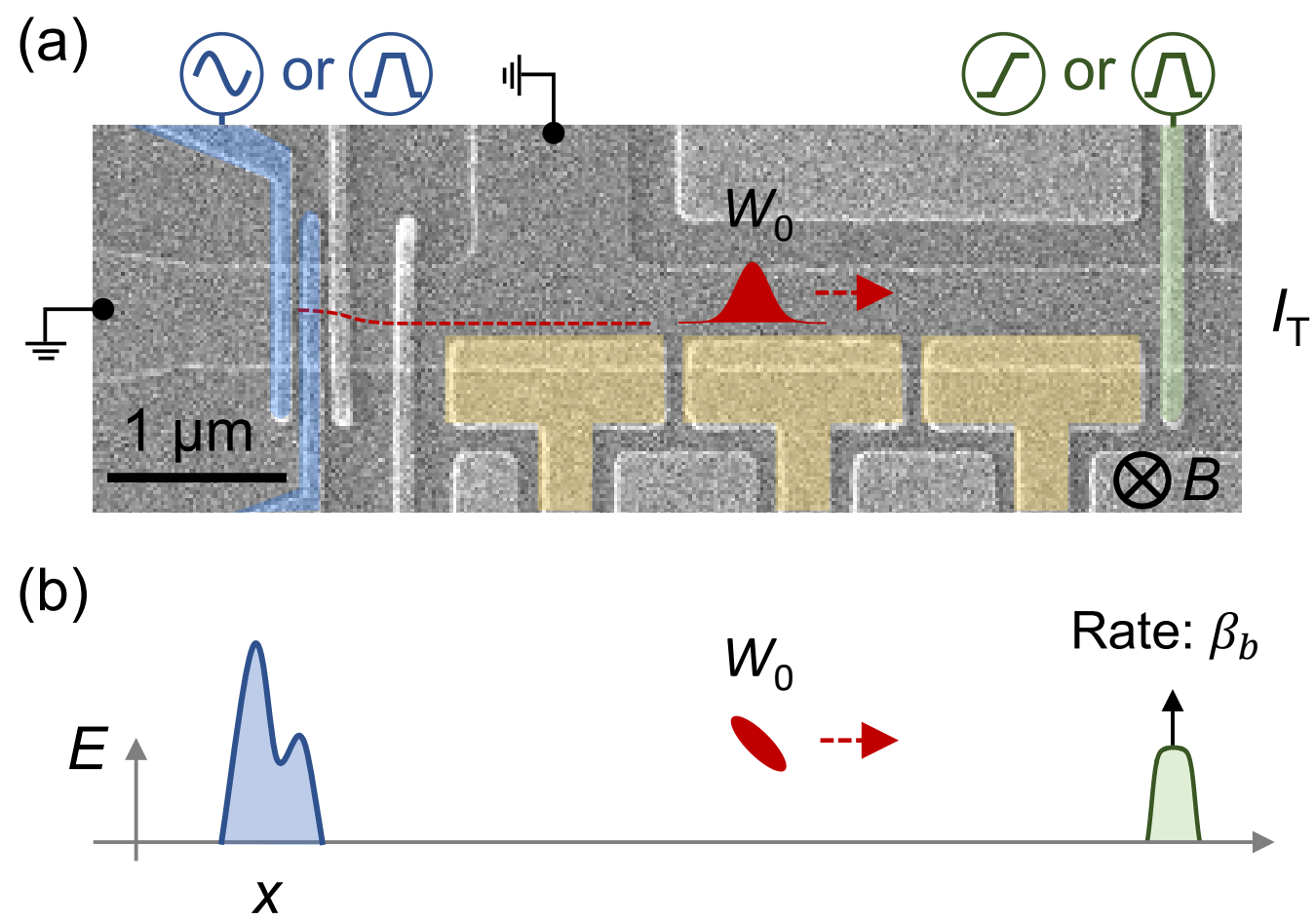

Figure 1. (a) Experimental setup. A quantum dot, formed by the two gate voltages (blue region), is temporally modulated by applying a sinusoidal or trapezoidal waveform on the left gate, producing single electrons periodically. $W_0$ represents the Wigner distribution of the emitted electrons. To perform time-resolved measurements, the detector potential barrier (green region) is also modulated periodically. (b) Schematic representation of the setup in position and energy space.

We present our experimental setup, which utilizes a 2DEG formed in a GaAs/AlGaAs heterostructure (see Supplementary Material, Sec. S1 for details). The setup is shown in Fig. 1(a), with the corresponding potential profiles along the channel in Fig. 1(b). For single-electron emission (excitation far above the Fermi energy), we modulate a quantum dot over time by controlling the voltage on the gate that defines the quantum dot (color blue in Fig. 1). We use either sinusoidal or trapezoidal waveforms at a pumping frequency $f_p$. The pumping process emits single electrons with emission energies ~100 meV above the Fermi energy [21]. The emitted electrons, characterized by the phase-space distribution $W_0$, propagate through a quantum Hall channel under a perpendicular magnetic field of $B = 12$ T, which corresponds to the filling factor 1. Side gates [yellow regions in Fig. 1(a)] with proper gate voltages are introduced to suppress the emission of longitudinal optical (LO) phonons during electron propagation through the quantum Hall channel [27]. For time-resolved measurements, sinusoidal or linearly ramping voltage waveform, synchronous with pumping, is applied to the detector barrier [formed by the green gate in Fig. 1(a)], and the current $(I_T)$ transmitted through the barrier is measured.

For the sinusoidal voltage waveform, we apply $V_b(t) = V_{\mathrm{AC}} \sin 2\pi f_b(t - t_d) + V_{\mathrm{DC}}$, where $t_d$ denotes the time delay relative to the pumping signal. In our experimental regime, the AC period $1/f_b$ ($\sim 4$ ns for $f_b = 240$ MHz) is much larger than both the temporal width of incident (emitted) electrons ($\leq 30$ ps) [23] and the traversal time of electrons through the barrier ($\leq 10$ ps) [28]. Under these conditions, the rate of change $\beta_b$ in the barrier height over time, around the arrival time of electrons at the barrier, can be described as linear given by $\beta_b \simeq -\alpha_b \partial V_b / \partial t|_{t=\langle t \rangle_b}$. Here, $\langle t \rangle_b$ is the mean arrival time of electrons at the barrier, and $\alpha_b$ is the conversion factor from gate voltage to barrier height. For the sinusoidal voltage, the rate depends on both the AC amplitude and the time delay $t_d$. The linearly ramping voltage waveform is obtained using a hyperbolic tangent function $V_b(t) = V_{\mathrm{AC}} \tanh[A \sin[2\pi f_b(t - t_d)]] + V_{\mathrm{DC}}$ as in Ref. [23]. In this waveform, the parameter $A$ controls the slope of the temporal profile and thus modulates the rate $\beta_b$.

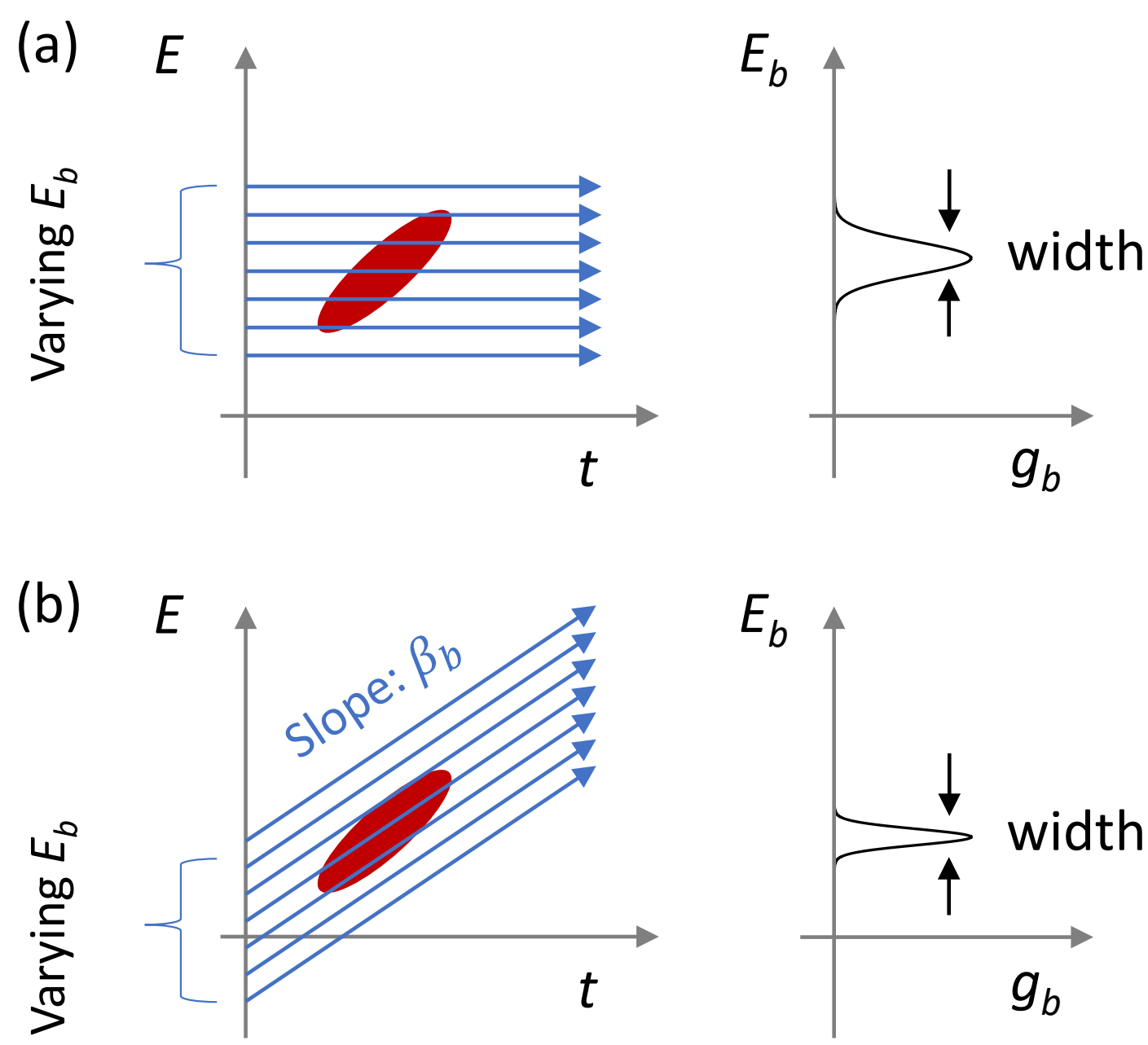

Figure 2. (a,b) Wigner distributions (red ovals) are shown in the energy-time phase space at the left panels. The projected distributions along the blue arrows are shown in right panels. The projected distribution is equivalent to the transconductance $g_b$ when the tunneling energy window of the potential barrier is negligible. They are shown for (a) a static barrier and (b) a barrier height linearly modulated with a rate $\beta_b$.

High-energy electrons have been characterized using transconductance measurements of the current $I_T$ through the detector barrier [21-23]. We define the (normalized) transconductance $g_b$ as

$$g_b \equiv -\frac{1}{ef_p}\frac{\partial I_T}{\partial E_b},$$

where $ef_p$ is the pumping current used to normalize the current, and $E_b \equiv -\alpha_b V_{\mathrm{DC}}$ denotes the DC component of the barrier height controlled by $V_{\mathrm{DC}}$. Under a linearized barrier height rate $\beta_b$, the transconductance $g_b$ is expressed as [24]

$$g_b = -\frac{\partial}{\partial E_b}[\int_{-\infty}^{\infty} dt \int_{-\infty}^{\infty} dE\, W_0(E,t)\, T(E - E_b - \beta_b t + \text{const})], \qquad (1)$$

where $W_0(E,t)$ is the Wigner distribution of the incident (emitted) electrons in the energy-time phase space. This Wigner distribution is the Weyl transform of a single-particle density matrix, representing the quasi-probability distribution in the phase-space [29]. This could be a low-purity mixed state, which may exhibit a classical-like distribution, or a pure state. Here, $T(E)$ is an energy-dependent transmission probability of monochromatic electrons of energy $E$ through the detector barrier, and "const" denotes a factor independent of $E$ and $t$.

For a static potential barrier ($\beta_b = 0$), the barrier acts as an energy filter [21], allowing a part of electrons whose energy exceeds the barrier height (treating $T(E)$ as a step function). In this case, Eq. (1) approximates the projection of the incident $W_0$ onto the energy space, yielding the energy distribution of electrons as shown in Fig. 2(a). For a finite $\beta_b$, Eq. (1) approximates the projection of $W_0$ onto the energy space along tilted lines [23-25,28] with slope $\beta_b$ as shown in Fig. 2(b).

We examine the variance in the barrier height (corresponding to the energy scale) of the transconductance $g_b$, $\langle \Delta E_b^2 \rangle_g \equiv \langle E_b^2 \rangle_g - \langle E_b \rangle_g^2$ for a given rate $\beta_b$, where $\langle E_b^k \rangle_g \equiv \int_{-\infty}^{\infty} dE_b\, E_b^k\, g_b$. This variance can be computed as (see Supplementary Material, Sec. S2),

$$\langle \Delta E_b^2 \rangle_g = \langle \Delta E^2 \rangle_0 + \Delta_b^2 - 2\beta_b \langle \Delta E \Delta t \rangle_0 + \beta_b^2 \langle \Delta t^2 \rangle_0, \qquad (2)$$

where $\langle O \rangle_0 \equiv \int_{-\infty}^{\infty} dE\, dt\, O\, W_0$, and $\Delta_b$ represents the energy width (standard deviation) of $\partial T/\partial E$. This width represents the energy window in which the transmission amplitude $T$ varies from 0 to 1. When the window is narrower than the energy spread of the incoming electrons, the barrier acts as the energy-time filter, as shown in Fig. 2.

The variance in barrier height in Eq. (2) at $\beta_b = 0$ approximates the energy variance $\langle \Delta E^2 \rangle_0$ of the incident electrons[21] under the condition that $\Delta_b^2$ is sufficiently small than $\langle \Delta E^2 \rangle_0$. We neglect this term $\Delta_b^2$, as the condition is assumed to be satisfied in experiments [30,31]. A finite value of $\Delta_b$ will introduce an additional deviation in the estimation of $\langle \Delta E^2 \rangle_0$. Independent determination of $\Delta_b^2$ and $\langle \Delta E^2 \rangle_0$ would be useful for quantifying errors by the detector barrier, for instance using multiple barriers as in Ref. [30]. The coefficient of $\beta_b^2$ corresponds to the temporal variance $\langle \Delta t^2 \rangle_0$, while the coefficient of $-2\beta_b$ corresponds to the energy-time covariance $\langle \Delta E \Delta t \rangle_0$. Here, this extraction does not depend on the detailed shape of $\partial T / \partial E$ and the constant term in Eq. (1). Therefore, by obtaining Eq. (2) experimentally and analyzing the resulting quadratic curve as a function of $\beta_b$, we can extract the components of the covariance matrix

$$\Sigma_0 = \begin{pmatrix} \langle \Delta E^2 \rangle_0 & \langle \Delta E \Delta t \rangle_0 \\ \langle \Delta E \Delta t \rangle_0 & \langle \Delta t^2 \rangle_0 \end{pmatrix}. \tag{3}$$

In the following, we demonstrate how these components can be determined experimentally.

Figure 3(a) shows the transconductance $g_b$ as a function of $t_d$ and $E_b$, measured while applying a sinusoidal waveform to the detector barrier with a frequency $f_b = 240$ MHz and amplitude $V_{\mathrm{AC}} = 82$ mV (see Supplementary Material, Sec. S3 for pumping conditions). Under these conditions, $\beta_b$ ranges from $-0.07$ to $0.07$ meV/ps. The higher-energy curve is produced by electrons that do not experience LO phonon emission during propagation on the way to the detector, while the lower-energy curve corresponds to electrons that emit one LO phonon [21,27]. The higher-energy curve is used to characterize the state of the incident electrons. By identifying the gate voltage that separates the two curves, the conversion factor $\alpha_b = 0.53 - 0.54$ eV/V is obtained [32] (see Supplementary Material, Sec. S4). It is observed that varying $t_d$ alters the barrier height rate $\beta_b$ (i.e., the slope of the energy-time curve) [21], and consequently changes the variance of the barrier height due to the time-energy filtering effect in Fig. 2 [21]. To confirm consistency with the results from linear ramps, we also apply hyperbolic tangent waveforms with various $\beta_b$ to the detector barrier at $f_b = 2.4$ GHz. $g_b$ for $\beta_b = 0.023$ meV/ps is shown in Fig. 3(b).

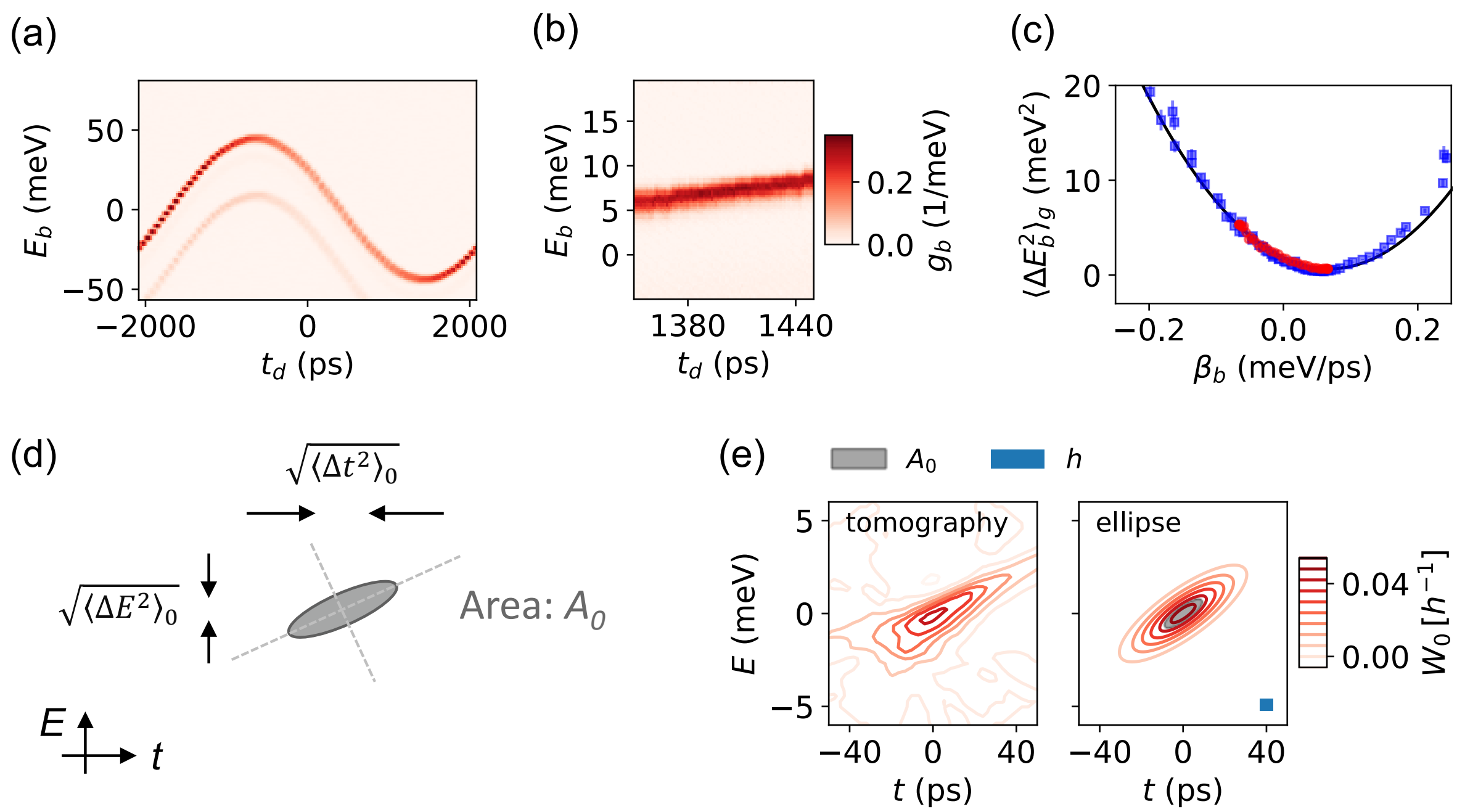

Figure 3. (a,b) Transconductance $g_b$ as a function of $t_d$ and $E_b$, measured with (a) a sinusoidal and (b) a linearly ramping waveform applied to the detector barrier. In both cases, a sinusoidal waveform with a pumping frequrncy of $f_p = 240$ MHz is used for electron emission. In (a), $V_{\mathrm{AC}} = 82$ mV and $f_b = 240$ MHz are used. In (b), $g_b$ is shown for $\beta_b = 0.023$ meV/ps with $f_b = 2.4$ GHz. (c) The barrier height variance $\langle \Delta E_b^2 \rangle_g$ obtained from (a) is shown as red circles, fitted to Eq. (2) (black solid curve). $\langle \Delta E_b^2 \rangle_g$ obtained from $g_b$ with the linearly ramping waveforms [as in (b)] is shown as blue squares. (d)

Energy width $\sqrt{\langle \Delta E^2 \rangle_0}$, time width $\sqrt{\langle \Delta t^2 \rangle_0}$, and the effective phase-space area $A_0$ (gray region). (e) Wigner distributions $W_0$ reconstructed from tomographic measurements (left) and derived from the covariance matrix $\Sigma_0$ assuming elliptical distributions (right). The gray region indicates the area computed from $\Sigma_0$. For comparison, the area corresponding to Planck constant $h = 2\pi\hbar$ is shown as the blue rectangles, while that of $\hbar/2$ is too small to be visible at this scale.

From the transconductance $g_b$ in Figs. 3(a) and (b), we extract the variance $\langle \Delta E_b^2 \rangle_g$ as a function of $\beta_b$ in Fig. 3(c) (see Supplementary Material, Sec. S4). Results for the sinusoidal waveform [Fig. 3(a)] are represented as red circles with error bars derived from the fitting process (though they are not visible due to their small size). These data are well-fitted to Eq. (2) (black solid curve), from which we extract $\langle \Delta t^2 \rangle_0 = (15.9 \pm 0.3 \text{ ps})^2$, $\langle \Delta E \Delta t \rangle_0 = 17.1 \pm 0.3 \text{ meV} \cdot \text{ps}$, and $\langle \Delta E^2 \rangle_0 \simeq (1.35 \pm 0.05 \text{ meV})^2$. These values align with the ranges reported in previous studies [22,23]. The positive sign of the covariance reflects that, during emission of electrons from the quantum dot pump, later-ejected parts have higher energy than earlier-ejected parts [23,25]. Results from the linearly ramping waveforms [blue squares with error bars in Fig. 3(c)] show good agreement with those from the sinusoidal waveform (see Supplementary Material, Sec. S6 for results in other samples).

The uncertainties are obtained by considering the agreement between the data and Eq. (2), together with the error bars of the gate voltage to energy conversion factor and the error by a finite barrier height (gate voltage) interval (see Supplementary Material, Secs. S4 and S5). Our results [red circles in Fig. 3(c)] obtained using a sinusoidal waveform, that capture the overall shape of the quadratic curve, provide reliable estimates (see Supplementary Material, Sec. S7). The measurement time using the sinusoidal waveform to obtain Fig. 3(c) is ~44 minutes (compared with 41 hours for tomographic methods in our experiment). This time could be further reduced by employing sparse sampling along the time delay $t_d$ domain and by measuring only the necessary portion of the data. We estimate that the measurement time can be optimized to ~7.4 minutes by using ~5550 data points with preliminary measurements (see Supplementary Material, Sec. S8).

We now analyze a property derived from the covariance matrix $\Sigma_0$ and define the effective phase-space area $A_0$ as

$$A_0 \equiv \sqrt{\det \Sigma_0} = \sqrt{\langle \Delta E^2 \rangle_0 \langle \Delta t^2 \rangle_0 - \langle \Delta E \Delta t \rangle_0^2}. \tag{4}$$

This quantity corresponds to the uncertainty product along the principal (tilted) axes [dashed lines in Fig. 3(d)], which represent the effective phase-space area (gray region), as detailed later. When the covariance vanishes, this expression reduces to the standard uncertainty product along the energy and time axes.

To obtain Eq. (4), we first express the covariance matrix $\Sigma_0$ in dimensionless coordinates. The principal axes and the variances of $W_0$ along the axes can be obtained from the eigenvectors and eigenvalues of the covariance matrix, respectively (see Supplementary Material, Sec. S9). The principal axes refer to the directions where the covariance of $W_0$ vanishes. Hence, the product of standard deviations along the principal axes quantifies the spread of $W_0$ in the phase space. Here, the product of the eigenvalues equals the determinant of the matrix. Converting back to dimensionful units then yields $A_0$ of Eq. (4) [see Fig. 3(d)]. Here, we note that the defined area of Eq. (4) is invariant under the choice of scaling factors used for dimensionless coordinates. When the distribution exhibits negative values, as in Ref. [33], the definition of the effective area may become ambiguous. Nevertheless, since the negative region is typically small, $A_0$ still captures the overall spread of the distribution, and serves as a useful measure.

According to the Robertson-Schrodinger uncertainty relation [34,35], $A_0$ is bounded by $A_0 \geq \hbar/2$. Under the elliptical approximation of $W_0$, the purity $\gamma_0$ is given by $\gamma_0 = \hbar/(2A_0)$, see Supplementary Material, Sec. S10 for details. The equality $A_0 = \hbar/2$ corresponds to a pure state of $\gamma_0 = 1$, while $A_0 > \hbar/2$ corresponds to a mixed state of $\gamma_0 < 1$. Therefore, comparing $A_0$ with $\hbar/2$ provides a convenient measure of the mixedness. We show later that the elliptical approximation based on the covariance matrix $\Sigma_0$, along with the effective area defined in Eq. (4), qualitatively captures the overall shape of the distribution reconstructed from the tomographic measurements. Representing the elliptical $W_0$ as a density matrix $\rho_0$ of the form $\rho_0 = \sum_{i=1}^{\infty} p_i \, |\psi_i\rangle\langle\psi_i|$, where $|\psi_i\rangle$ are normalized pure states with weights $p_i$, we obtain that the largest weight is $\max p_i = 2\hbar/(2A_0 + \hbar)$ (see Supplementary Material, Sec. S10). Therefore, obtaining $A_0$ allows us to estimate this weight. This result is consistent with Ref. [23], where the maximum weight is estimated from the maximum probability of the reconstructed Wigner distribution, giving $\max p_i \simeq \hbar/A_0 = h \max W_0$ in the regime $A_0 \gg \hbar/2$. Our method, in contrast, requires only a sinusoidal wave to obtain $A_0$ (or equivalently $\Sigma_0$) for the estimation.

The left panel in Fig. 3(e) shows $W_0$ reconstructed from tomographic measurements, while the right panel shows that obtained from $\Sigma_0$ under elliptical approximation. We find that the two distributions show qualitative agreement. The effective area [Eq. (4)] computed from $\Sigma_0$ is represented by the gray region in the right panel (see Supplementary Material, Sec. S9). The extracted area is $A_0 \sim (20 \pm 2)\hbar$, which is significantly larger than the quantum limit $\hbar/2$ and consistent with previous reports on low purity [23]. This may be attributed to voltage noise or phase jitter [21]. See Supplementary Material, Sec. S5 for computing an uncertainty of $A_0$. Some discrepancies between the two distributions are observed; however, it is difficult to distinguish them from errors in our experiment, such as a limited range of tomographic angle (see Supplementary Material, Sec. S11) and experimental noises for high ramping rate.

We then apply our method to study how the electron state varies under different pumping condition [23,33,36]. Specifically, we emit electrons using a trapezoidal waveform applied to the quantum dot and vary the pumping frequency $f_p$ (see Supplementary Material, Sec. S3 for pumping conditions). Figure 4(a) shows the barrier height variance $\langle \Delta E_b^2 \rangle_g$ obtained at $f_p =$ 120 MHz (blue square) and $f_p = 240$ MHz (red square). Following the same procedure in Fig. 3, we extract $\langle \Delta t^2 \rangle_0 = (16.8 \pm 0.3 \text{ ps})^2$, $\langle \Delta E \Delta t \rangle_0 = 23.1 \pm 0.6 \text{ meV} \cdot \text{ps}$, and $\langle \Delta E^2 \rangle_0 \simeq (1.48 \pm 0.02 \text{ meV})^2$ for $f_p = 120$ MHz. For $f_p = 240$ MHz, we obtain $\langle \Delta t^2 \rangle_0 = (11.8 \pm 0.1 \text{ ps})^2$, $\langle \Delta E \Delta t \rangle_0 = 18.6 \pm 0.3 \text{ meV} \cdot \text{ps}$, and $\langle \Delta E^2 \rangle_0 \simeq (1.72 \pm 0.03 \text{ meV})^2$.

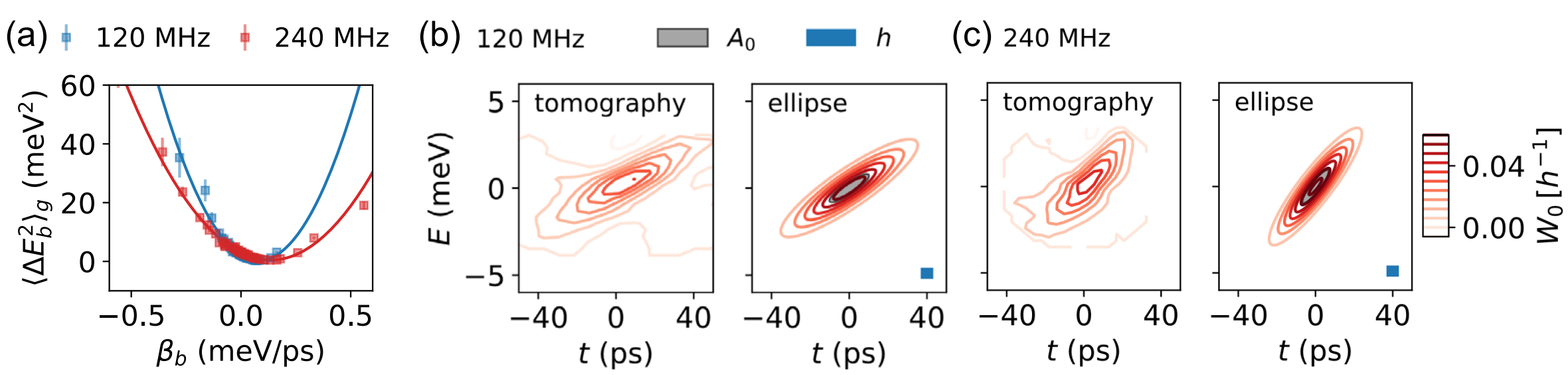

Figure 4. (a) Barrier height variance $\langle \Delta E_b^2 \rangle_g$ as a function of $\beta_b$, obtained from $g_b$ with linearly ramping waveforms applied to the detector barrier. Results are shown in blue for $f_p = 120$ MHz and in red for $f_p =$ 240 MHz. Solid lines represent fits to Eq. (2). (b, c) Wigner distributions $W_0$ reconstructed from tomographic measurements (left) and from the covariance matrix $\Sigma_0$ under elliptical approximation (right),

for (b) $f_p = 120$ MHz and (c) $f_p = 240$ MHz. The area of the gray region indicates the effective area $A_0$, computed from the covariance matrix.

Figures 4(b) and (c) show the Wigner distributions $W_0$, reconstructed from tomographic measurements (left panels) and obtained from the covariance matrix $\Sigma_0$ under the elliptical approximation (right panels). The two distributions exhibit qualitative agreement. The areas $A_0$ are $\sim (14 \pm 3)\hbar$ and $(12 \pm 2)\hbar$ for $f_p = 120$ MHz and $240$ MHz, respectively, much larger than $\hbar/2$. As the pumping frequency $f_p$ increases, electron emission from the quantum dot occurs over a shorter timescale, leading to a smaller temporal variance $\langle \Delta t^2 \rangle_0$ [23,33]. At the same time, due to rapid energy variation during emission, the energy variance $\langle \Delta E^2 \rangle_0$ becomes large. Our results confirm these trends, demonstrating the broad operability of our measurement method across varying quantum dot emission conditions.

In summary, we have developed a measurement method to extract the covariance matrix of single electrons, capturing their energy variance, temporal variance, and energy-time covariance by temporally modulating the detector potential barrier. The extraction is enabled by measuring the energy variance in transconductance. This method works for both sinusoidal and linearly ramping (hyperbolic tangent) voltage waveforms, showing consistency across the two waveforms. The sinusoidal waveform proves efficient and precise within our framework, as it requires shorter measurement times and avoids the need for high-frequency waveform synthesis.

From the covariance matrix, we define the effective phase-space area and construct an elliptical distribution. The results show qualitative agreements with that obtained from the tomographic measurements. Our results confirm that the extracted parameters vary under different pumping conditions in an expected manner. We anticipate that the proposed measurement method will be useful for efficiently examining quantum states and will be valuable in electron quantum optics and sensing.

This work was supported by the National Research Foundation funded by the Korean Government (Grant Nos. RS-2023-00207732, RS-2021-NR059826, RS-2023-00256050, RS-2023-00283291, RS-2024-00463743, and RS-2024-00443714). This research was also partially supported by Development of quantum-based measurement technologies funded by Korea Research Institute of Standards and Science (KRISS-2025-GP2025-0010).

**Author Declarations**

**Conflict of interest**

The authors have no conflicts to disclose.

**Author contributions**

**Wanki Park:** Conceptualization (lead); Formal analysis (lead); Investigation (lead); Writing–original draft (lead); Writing–review & editing (lead).

**Chanuk Yang:** Data curation (equal); Investigation (equal); Methodology (equal).

**Young-Seok Ghee:** Methodology (equal); Resources (equal).

**Hyung Kook Choi:** Methodology (supporting); Resources (supporting).

**Bum-Kyu Kim:** Methodology (equal); Resources (equal).

**Myung-Ho Bae:** Supervision (lead); Funding acquisition (lead); Data curation (equal); Investigation (equal); Project administration (lead); Writing–review & editing (supporting).

**Data availability**

The data that support the findings of this study are available from the corresponding authors upon reasonable request.

Supplementary Material for

# Method for rapid estimation of the energy-time covariance matrix of single electrons

Wanki Park[1*†], Chanuk Yang[2,3†], Young-Seok Ghee[1], Hyung Kook Choi[3], Bum-Kyu Kim[1] and Myung-Ho Bae[1,4*]

[1]Korea Research Institute of Standards and Science, Daejeon 34113, Republic of Korea

[2]Jeonbuk National University G-LAMP Project Group, Jeonbuk National University, Jeonju 54896, Republic of Korea

[3]Department of Physics, Jeonbuk National University, Jeonju 54896, Republic of Korea

[4]KAIST Graduate School of Quantum Science and Technology, Korea Advanced Institute of Science and Technology, Daejeon 34141, Republic of Korea

[†]W.P. and C.Y. contributed equally to this work.

*e-mail: wanki931@gmail.com, mhbae@kriss.re.kr

## S1. Sample fabrication and measurement

The GaAs/AlGaAs device is fabricated with the top-gate electrodes, as shown in Fig. 1(a) of the main text. A 2DEG is generated at a depth of 70 nm beneath the surface. The carrier density and mobility are ~$2.5\times10^{11}$ $cm^{-2}$ and ~70 $m^2V^{-1}s^{-1}$, respectively. The device is cooled down to 420 mK. The electric current is measured at a terminal positioned across the barrier using an ammeter, which consists of an *I*/*V* converter-type current amplifier with a gain of $10^9$ and a digital voltmeter. A Tektronix AWG7122C arbitrary waveform generator (AWG) is used to generate the waveforms for both pumping and detection in Figs. 3 and 4 of the main text. This AWG is also used for tomographic measurements in Sec. S11.

## S2. Derivation of Eq. (2) of the main text

We present a detailed derivation of Eq. (2) from Eq. (1), introduced in the main text. The transconductance of Eq. (1) can be expressed in convolution form as

$$g_b = -\frac{\partial}{\partial E_b}\int_{-\infty}^{\infty} dt \int_{-\infty}^{\infty} dE\, W_0(E,t)T(E - E_b - \beta_b t + \text{const}) \tag{S1}$$

$$= \int_{-\infty}^{\infty} dE\, \underbrace{\int_{-\infty}^{\infty} dt\, W_0(E + \beta_b t - \text{const}, t)}_{f(E)} \underbrace{\left[-\frac{\partial}{\partial E_b} T(E - E_b)\right]}_{j(E_b - E)} \tag{S2}$$

$$= \int_{-\infty}^{\infty} dE\, f(E) j(E_b - E). \tag{S3}$$

We note that the integral of either function $f$ or $j$ over the energy (barrier height) is 1. From the convolutional form of $g_b$ [Eq. (S3)], the variance in the barrier height is given by the sum of the variance of the functions $f$ and $j$,

$$\langle \Delta E_b^2 \rangle_g \equiv \int_{-\infty}^{\infty} dE_b\, E_b^2 g_b - \left[\int_{-\infty}^{\infty} dE_b\, E_b g_b\right]^2 \tag{S4}$$

$$= \int_{-\infty}^{\infty} dE\, E^2 f - \left[\int_{-\infty}^{\infty} dE\, Ef\right]^2 + \int_{-\infty}^{\infty} dE_b\, E_b^2\, j - \left[\int_{-\infty}^{\infty} dE_b\, E_b\, j\right]^2. \tag{S5}$$

From the definition of $f$ in Eq. (S2), we obtain

$$\int_{-\infty}^{\infty} dE\, E^2 f - \left[\int_{-\infty}^{\infty} dE\, Ef\right]^2 = \langle \Delta E^2 \rangle_0 - 2\beta_b \langle \Delta E \Delta t \rangle_0 + \beta_b^2 \langle \Delta t^2 \rangle_0, \tag{S6}$$

where $\langle O \rangle_0 \equiv \int_{-\infty}^{\infty} dE\, dt\, O\, W_0$. Defining the square root of $\int_{-\infty}^{\infty} dE_b\, E_b^2\, j - \left[\int_{-\infty}^{\infty} dE_b\, E_b\, j\right]^2$ as $\Delta_b$, i.e., the energy width of $\partial T/\partial E$, and using Eqs. (S5) and (S6), Eq. (S4) becomes Eq. (2) in the main text.

**S3. Pumping conditions for Figs. 3 and 4 of the main text**

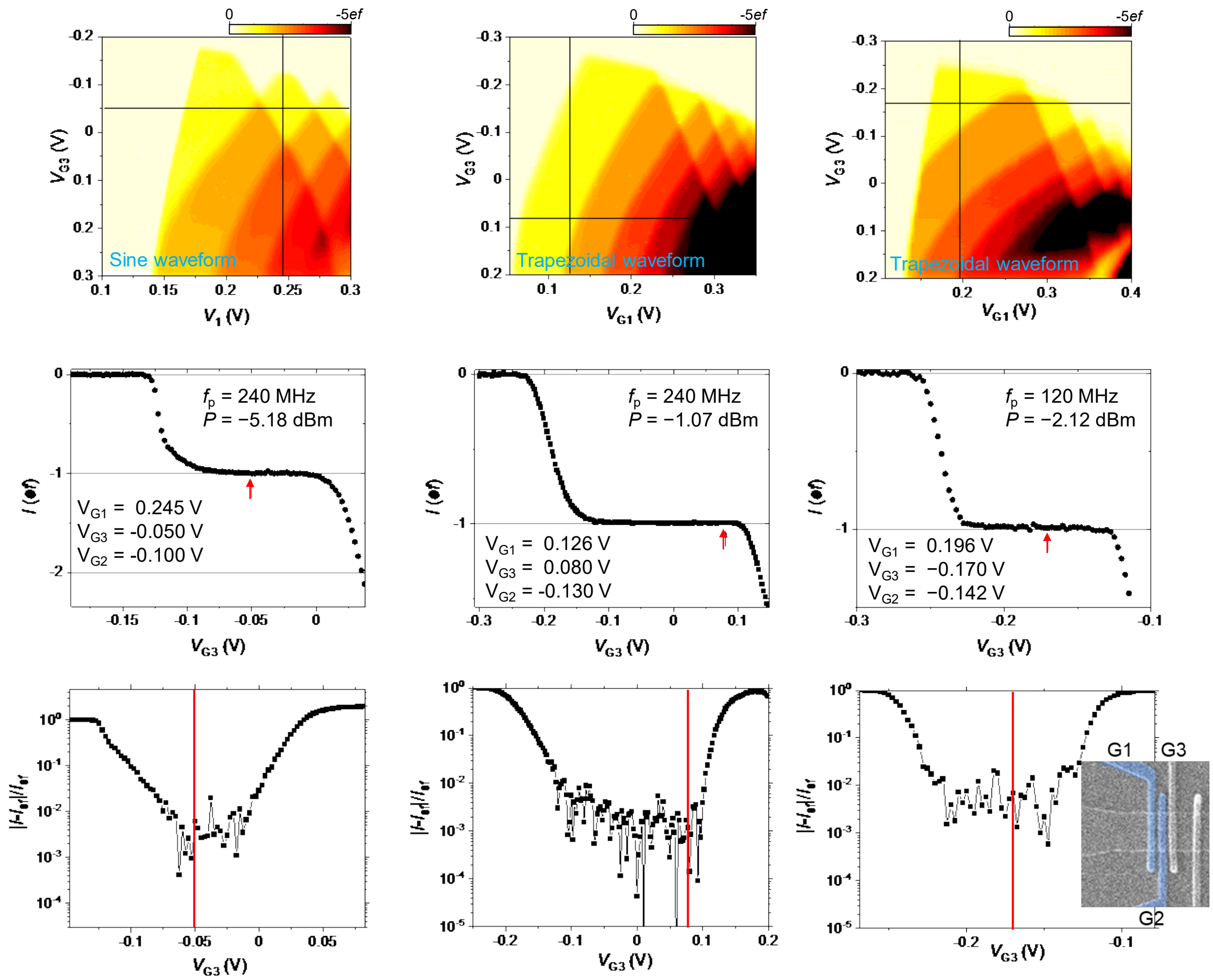

Figure S1. Pumping conditions for Figs. 3 (sine waveform, $f_p$ = 240 MHz) and 4 (trapezoidal waveforms, $f_p$ = 240 and 120 MHz) of the main text. Top panels of (a), (b) and (c): pump maps as functions of $V_{G3}$ and $V_{G1}$ for Fig. 3, for Fig. 4 at $f_p$ = 240 MHz and 120 MHz, respectively. The corresponding $V_{G2}$ values were −0.100 V, −0.130 V and −0.142 V, respectively. Here, gates G1, G2 and G3 correspond to the left blue gate, right blue gate, and the adjacent gate, respectively, as shown in the inset of the bottom panel of (c). The experimental pumping condition in the main text is indicated by a crosshair in each map. The middle panels of (a), (b) and (c): Current ($I$) as a function of $V_{G3}$ along the vertical line in the corresponding pump map, with the experimental condition marked by a red arrow in each plot. The bottom panels of (a), (b) and (c): the relative deviation from $ef$, defined as (=$\left|\left(I - I_{ef}\right)\right|/I_{ef}$), as a function of $V_{G3}$. A vertical line in each figure indicates the $V_{G3}$ condition used in the experiments.

## S4. Error analysis of the gate conversion factor

Here, we evaluate the uncertainty of the gate-voltage-to-energy conversion factor (alpha factor) derived from the LO phonon emission [1], which contributes to the uncertainties in the extracted parameters. Figure S2 shows the transconductance as a function of the time delay $t_d$ and gate voltage $V_b$. The white solid line represents a sinusoidal fit that closely matches the experimental data as shown by the overlap. The conversion factor is determined by comparing the gate-voltage spacing between adjacent peaks—with and without LO phonon emission—to the LO phonon energy of 36 meV. From the peak separation labeled "A" and "B" in the figure, we obtain the gate conversion factors $\alpha$ of 0.54 and 0.53 eV/V, respectively. This corresponds to an uncertainty $2\%$. Similarly, applying the same procedure to Fig. S3 and Fig. S4 (see Sec. S6) yields uncertainties of 4% and 13%, respectively.

We incorporate the uncertainty of the gate conversion factor $\alpha$ into the uncertainties of the extracted parameters. Equation (2) of the main text can be expressed in terms of experimentally controlled gate-voltage parameters as

$$\alpha^2\langle\Delta V^2\rangle_g = \langle\Delta E^2\rangle_0 + \Delta_b^2 + 2\alpha\dot{V}\langle\Delta E\Delta t\rangle_0 + \alpha^2\dot{V}^2\langle\Delta t^2\rangle_0. \qquad \text{(S7)}$$

Here, $\langle\Delta V^2\rangle_g$ is the variance of the transconductance $g$ with respect to the DC gate voltage, and $\dot{V}$ is the rate of the time-dependent gate voltage.

To evaluate the effect of a small shift $\alpha \to \alpha \pm \delta\alpha$ (with $\delta\alpha \ll \alpha$), we examine the resulting variations in the parameters. By performing a Taylor expansion in $\delta\alpha/\alpha$, we find that the shift corresponds to $\langle\Delta E^2\rangle_0 \to \langle\Delta E^2\rangle_0(1 \mp 2\delta\alpha/\alpha)$ and $\langle\Delta E\Delta t\rangle_0 \to \langle\Delta E\Delta t\rangle_0(1 \mp \delta\alpha/\alpha)$, while $\langle \Delta t^2 \rangle_0$ remains unchanged. Here, the positive shift of $\alpha$ makes underestimation on $\langle \Delta E^2 \rangle_0$ and $\langle \Delta E \Delta t \rangle_0$. These variations indicate that the uncertainties of the extracted parameters are governed by $\delta\alpha/\alpha$. We therefore include these contributions in the total uncertainty of the extracted parameters, as discussed in Sec. S5.

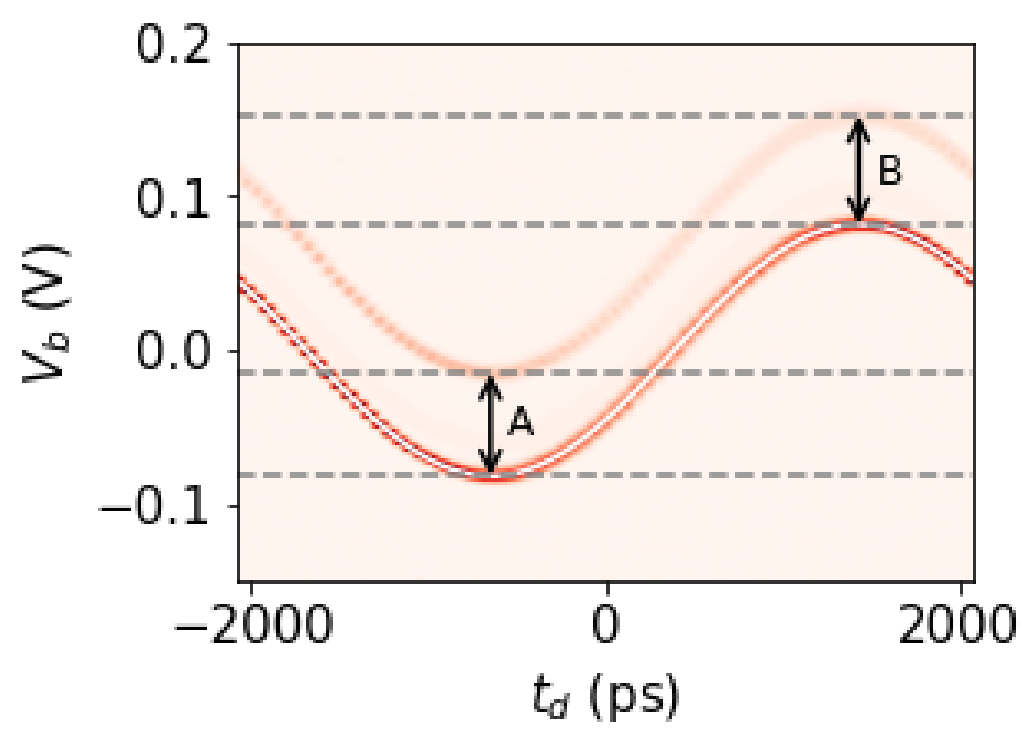

Figure S2. Transconductance as a function of $t_d$ and $V_b$, shown as an extended version of Fig. 3(a) in the main text but plotted with $V_b$ instead of the barrier height $E_b$. The white solid line (overlapped with data) shows a sinusoidal fit, and the peak separations “A” and “B” are indicated.

## S5. Parameter extraction from the transconductance and error analysis

The rate $\beta_b$ is obtained from the slope of the finite-value region in the transconductance. The value of $\langle \Delta E_b^2 \rangle_g$ is obtained by fitting a Gaussian curve to the cross-sectional profile of the transconductance $g_b$ along the barrier height $E_b$, as the Gaussian function accurately describes the shape of $g_b$ in our experiment. Our method can be also applied to electron wave packets of arbitrary shape by manually computing the variance through integration over finite regions of the transconductance. The extraction is performed using Lmfit, an open-source software package in Python [2]. It extracts $\langle \Delta E_b^2 \rangle_g$ via the least-squares method by minimizing the chi-square value, which quantifies the weighted sum of squared differences between the experimental data and the model. The errors of the width in extractions are shown as error bars in Figs. 3(c), 4(a), S4, and S6.

We extract the energy variance, temporal variance, and energy-time covariance by identifying the parameters using the least square method that minimizes the chi-square $\chi^2$, using Lmfit. The extrinsic uncertainties of the extracted parameters between the formula and experimental data are determined by the range of values, where the change in chi-square becomes significant. We also account for extrinsic uncertainties arising from the gate conversion factor (see Sec. S4 for details). The uncertainty arising from the finite barrier height (gate voltage) interval is also included, as analyzed in Sec. S8. The total uncertainties in the extracted parameters are obtained by summing these contributions in quadrature, as presented in the main text and Sec. S6.

To compute the uncertainty $\delta A_0$ of the effective phase-space area $A_0$ [Eq. (4) of the main text], assuming that the energy variance, time variance, and energy-time covariance are statistically independent, the uncertainty is given by

$$\delta A_0 = \frac{1}{2A}\sqrt{(\delta\langle \Delta E^2 \rangle_0)^2 \langle \Delta t^2 \rangle_0^2 + \langle \Delta E^2 \rangle_0^2 (\delta\langle \Delta t^2 \rangle_0)^2 + 4\langle \Delta E \Delta t \rangle_0^2 (\delta\langle \Delta E \Delta t \rangle_0)^2}, \tag{S8}$$

where $\delta\langle \Delta E^2 \rangle_0$, $\delta\langle \Delta t^2 \rangle_0$, and $\delta\langle \Delta E \Delta t \rangle_0$ are the uncertainties in $\langle \Delta E^2 \rangle_0$, $\langle \Delta t^2 \rangle_0$, and $\langle \Delta E \Delta t \rangle_0$, respectively. By evaluating Eq. (S8) using the extracted parameters and their uncertainties, we extract the uncertainty $\delta A_0$, whose values are provided in the main text and Sec. S6.

The energy resolution (gate voltage interval converted in energy scale) used for the detector barrier is 0.5 meV for Fig. 3 of the main text, 0.15 meV for Fig. 4 of the main text, and 0.25 meV for Fig. S3. Although these values are smaller than the measured energy width, this may cause to the uncertainty in measured energy, whose exact relation depends on the experimental signal to noise.

## S6. Measurement for other samples

We demonstrate our measurement method on different sample and setups. We first demonstrate our method [Fig. S3] in the same setup in Fig. 1(a) of the main text, but with a higher detector frequency and in a different setup. We then perform our method in a different setup [Fig. S4] while modifying electron states by varying voltage conditions in the pump map. From the results, we show the robust operability of our measurement method in a different barrier frequency and in the samples where experimental noises are expected to be large.

Figure S3(a) shows the barrier height variance $\langle \Delta E_b^2 \rangle_g$ at a pumping frequency $f_p = 960$ MHz and a detecting barrier frequency $f_b = 960$ MHz, as in Fig. 3(c) of the main text. Sinusoidal waveforms are used for both pumping and detecting. The data is well-fitted to Eq. (2) of the main text (black solid curve). From this fit, we extract $\langle \Delta t^2 \rangle_0 = (12.9 \pm 0.3\ \text{ps})^2$, $\langle \Delta E \Delta t \rangle_0 = 16.4 \pm 0.5\ \text{meV} \cdot \text{ps}$, and $\langle \Delta E^2 \rangle_0 \simeq (2.62 \pm 0.05\ \text{meV})^2$. The effective phase space area follows $A_0 = (45 \pm 2)\hbar$. The extracted $\langle \Delta E^2 \rangle_0$ is much larger than that reported in previous studies [3,4] and the extracted $A_0$ is much larger than the value in the main text, indicating the presence of additional noise in this sample.

Assuming that generated electrons are described by a bivariate Gaussian distribution (see Sec. S10), the Wigner distribution of the generated electrons is visualized in Fig. S3(b). The detector frequency of $f_b = 960$ MHz is larger than 240 MHz used in Fig. 3 of the main text. This large $f_b$ allows for a wide range of $\beta_b$ for a given $V_{\mathrm{AC}}$. The period $1/f_b \sim 1000$ ps is sufficiently long compared to the temporal width ~13 ps of the incident electrons. Thus, in our experimental regime, the barrier height rate $\beta_b$ can be approximated as linear during the traversal of electrons through the detector, and Eq. (2) of the main text remains valid.

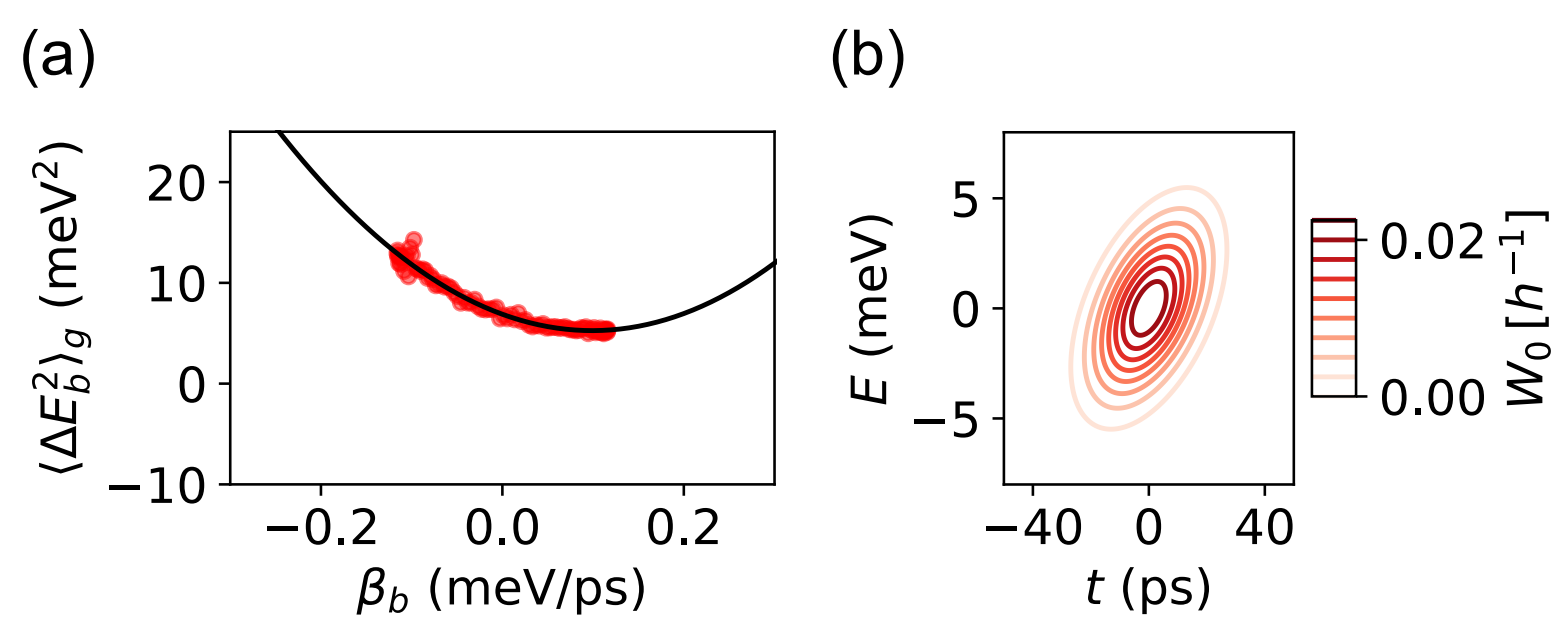

Figure S3. Results at a pumping frequency $f_p = 960$ MHz, detecting barrier frequency $f_b = 960$ MHz, $V_{\mathrm{AC}} = 40$ mV, and $\alpha = 0.47 \pm 0.02$ eV/V ($\alpha$ deduced from LO phonon emissions). (a) The barrier height variance $\langle \Delta E_b^2 \rangle_g$ obtained using a sinusoidal wave on the detector is shown as the red circles, fitted to Eq. (2) of the main text (the black solid curve). (b) A Wigner distribution $W_0$ of incident electrons is shown, assuming a bivariate Gaussian form described by an ellipse with the parameters obtained in (a).

Figure S4 shows results that are performed in a different setup (referred to the pump device 2 in this Supplemental Material), whose SEM image is shown as Fig. S4(a). The pump map of $dI/dV_{\mathrm{G3}}$ is shown in Fig. S4(b), where the points labeled P1, P2, and P3 (within the region of single-electron pumping) mark the conditions under which our method was implemented. Figure S4(c) displays transconductance $g_b$ for the three conditions in Fig. S4(b), measured by applying the sinusoidal waveform ($f = 480$ MHz) to the detector barrier. In Fig. S4(d), the barrier height variances extracted from the $g_b$ in Fig. S4(c) are shown as a function of the ramping rate $\beta_b$ as red circles, fitted to Eq. (2) (red solid curves) of the main text. To confirm consistency with different waveforms, the barrier height variances extracted using a linearly ramped (hyperbolic tangent) waveform are plotted as blue squares with the corresponding blue fitting curves. The extracted parameters of electrons obtained from these fits are summarized in Table S1. We find that the values extracted from the two waveforms show good agreement.

We discuss the results as the pumping condition approaches the emission line [the dotted line in Fig. S4(a)], where electrons are unable to escape from the quantum dot. As the pumping condition changes from P3 to P1, it takes longer time for tunneling in the emission process, and the temporal variance will increase while the energy variance decreases [3]. Our results follow this trend (Table S1); however, the parameters do not change significantly. This may be attributed to the fact that the pump device 2 of Fig. S4(a) is vulnerable to the environmental noise, which could significantly influence the electron states. The effective phase-space area, $A_0 \sim (70 \pm 20)\hbar$, obtained from the sinusoidal waveform, is much larger than the values presented in the main text. This supports that the electron states are predominantly affected by noise. The larger uncertainty in $A_0$ is attributed to fluctuations of the gate conversion factor (with an uncertainty 13%) in this device.

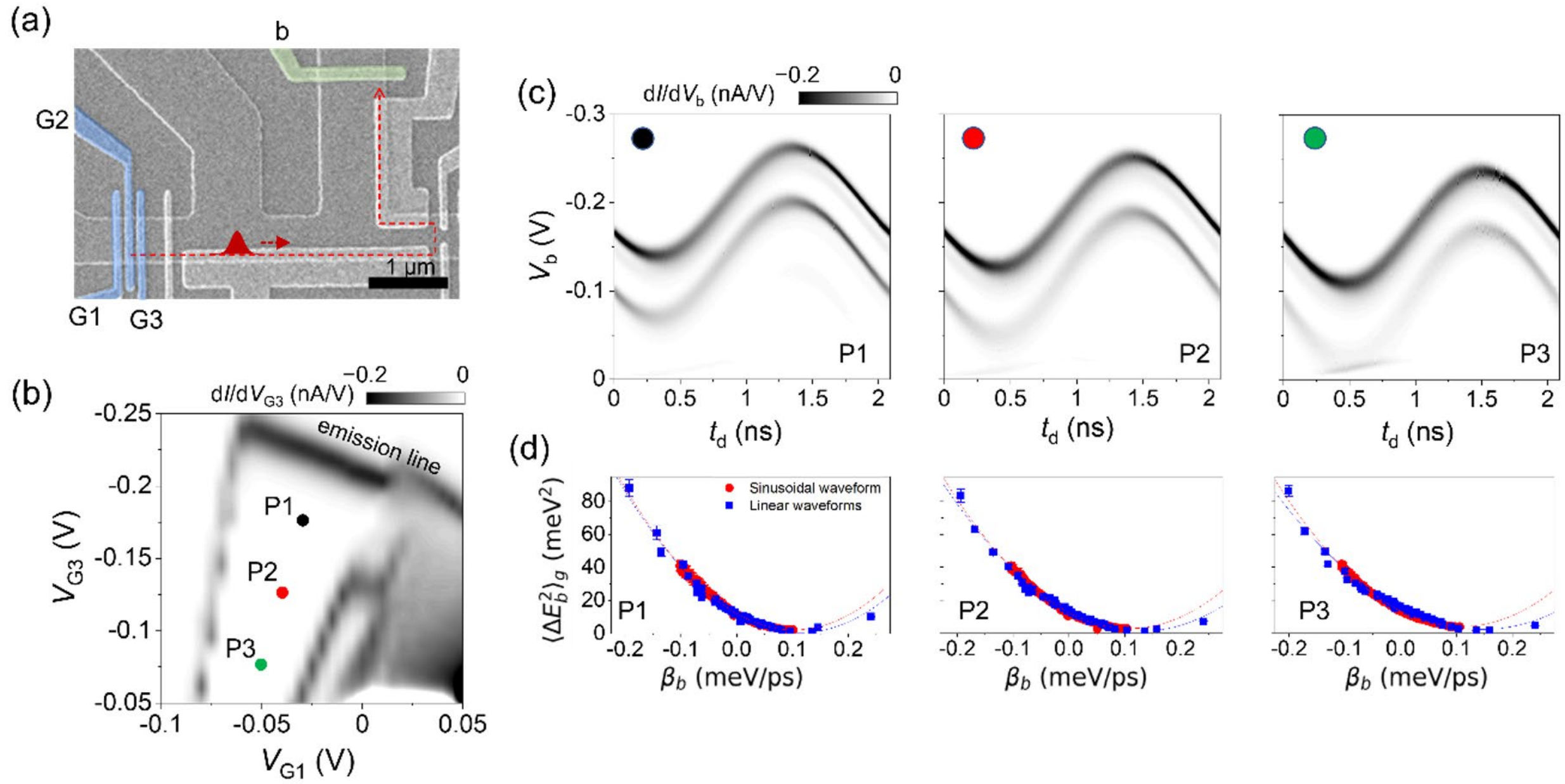

Figure S4. (a) SEM image of the pump device 2. (b) $dI/dV_{G3}$ map of the pump. The gate voltage conditions at the three locations (P1-P3) within single-electron pumping region are: ($V_1$, $V_3$) = (−0.03 V, −0.175 V), (−0.04 V, −0.125 V), and (−0.05 V, −0.075 V), respectively. The line where electrons cannot be emitted from the quantum dot at the emission stage is shown as the dotted line with a label 'emission line'. (c) Transconductance maps $g_b$ as a function of the time delay $t_d$ and the barrier gate voltage $V_b$, for the three locations. (d) Barrier height variances $\langle \Delta E_b^2 \rangle_g$ as a function of $\beta_b$ extracted from the $g_b$ maps in (c) are shown as red circles and fitted to Eq. (2) (red solid curve). Blue squares represent $\langle \Delta E_b^2 \rangle_g$ obtained using hyperbolic tangent waveforms.

| | Sinusoidal waveform | Linear waveforms |
|---|---|---|
| P1 | $(3.46 \pm 0.45)^2$ | $(3.41 \pm 0.45)^2$ |
| P2 | $(3.54 \pm 0.46)^2$ | $(3.60 \pm 0.47)^2$ |
| P3 | $(3.67 \pm 0.48)^2$ | $(3.89 \pm 0.51)^2$ |
| | **Sinusoidal waveform** | **Linear waveforms** |
| P1 | $(30.69 \pm 0.14)^2$ | $(29.54 \pm 0.55)^2$ |
| P2 | $(28.72 \pm 0.22)^2$ | $(26.75 \pm 0.52)^2$ |
| P3 | $(28.27 \pm 0.14)^2$ | $(24.66 \pm 0.70)^2$ |
| | **Sinusoidal waveform** | **Linear waveforms** |
| P1 | $(96.28 \pm 12.49)^2$ | $(98.86 \pm 13.03)^2$ |
| P2 | $(89.19 \pm 11.59)^2$ | $(91.88 \pm 12.08)^2$ |
| P3 | $(86.83 \pm 11.27)^2$ | $(88.29 \pm 11.71)^2$ |

Table S1. The energy variance $\langle \Delta E^2 \rangle_0$, temporal variance $\langle \Delta t^2 \rangle_0$, and energy-time covariance $\langle \Delta E \Delta t \rangle_0$ of electrons, extracted from the analysis of data in Fig. S4(d).

## S7. Range of barrier rate $\boldsymbol{\beta_b}$ for parameter extraction

In this section, we investigate the dependence of fitting results on a range of barrier ramping rate $\beta_b$. For this purpose, we crop the experimental data (red circles) in Fig. 3(c) of the main text and perform fits on subsets. We observe convergence in the extracted parameters as the $\beta_b$ range becomes sufficient to capture the overall shape of the quadratic curve [Eq. (2)]. We then show that the barrier height variance is consistent across two transconductance regions [Fig. S8] that share the same range of ramping speed $\beta$, indicating the robustness of our method.

A relevant parameter for the experimental data to capture the overall shape of the quadratic curve [see Fig. S5 for illustration] is the rate $\beta_b$ giving the minimum barrier height variance of $\langle \Delta E_b^2 \rangle_g$. From Eq. (2), the minimum is achieved at $\beta_b = \langle \Delta E \Delta t \rangle_0 / \langle \Delta t^2 \rangle_0$ (vertical red dotted line in Fig. S5). Another parameter is the rate $\beta_b = [\langle \Delta E^2 \rangle_0 / \langle \Delta \mathrm{t}^2 \rangle_0]^{1/2}$ (vertical green dotted line), at which the barrier height variance $\langle \Delta E_b^2 \rangle_g$ becomes larger than twice the initial energy variance $\langle \Delta E^2 \rangle_0$. The $\beta_b$ range covering the above relevant parameters would be a sufficient condition to obtain the parameters with accuracy.

Figure S6 shows $\langle \Delta E_b^2 \rangle_g$ for various $\beta_b$ ranges. The dashed lines represent fits based on Eq. (2) of the main text. The extracted values of $\langle \Delta E^2 \rangle_0$, $\langle \Delta t^2 \rangle_0$, and $\langle \Delta E \Delta t \rangle_0$, along with their associated uncertainties, are indicated by shaded regions. Figure S7 shows that, as the $\beta_b$ range increases, the extracted values of $\langle \Delta t^2 \rangle_0$ and $\langle \Delta E \Delta t \rangle_0$ converge to stable values when the maximum of the $\beta_b$ range becomes comparable to $\beta_b = \langle \Delta E \Delta t \rangle_0 / \langle \Delta t^2 \rangle_0$ (red dashed line) and the order of maximum of the $\beta_b$ range is comparable to $\beta_b = [\langle \Delta E^2 \rangle_0 / \langle \Delta t^2 \rangle_0]^{1/2}$ (blue dotted line). Although the $\beta_b$ range does not cover large values of $\beta_b$, the agreements of sinusoidal data with results of hyperbolic tangent [blue squares in Fig. 3(c) of the main text] for large range of $\beta_b$ support our arguments.

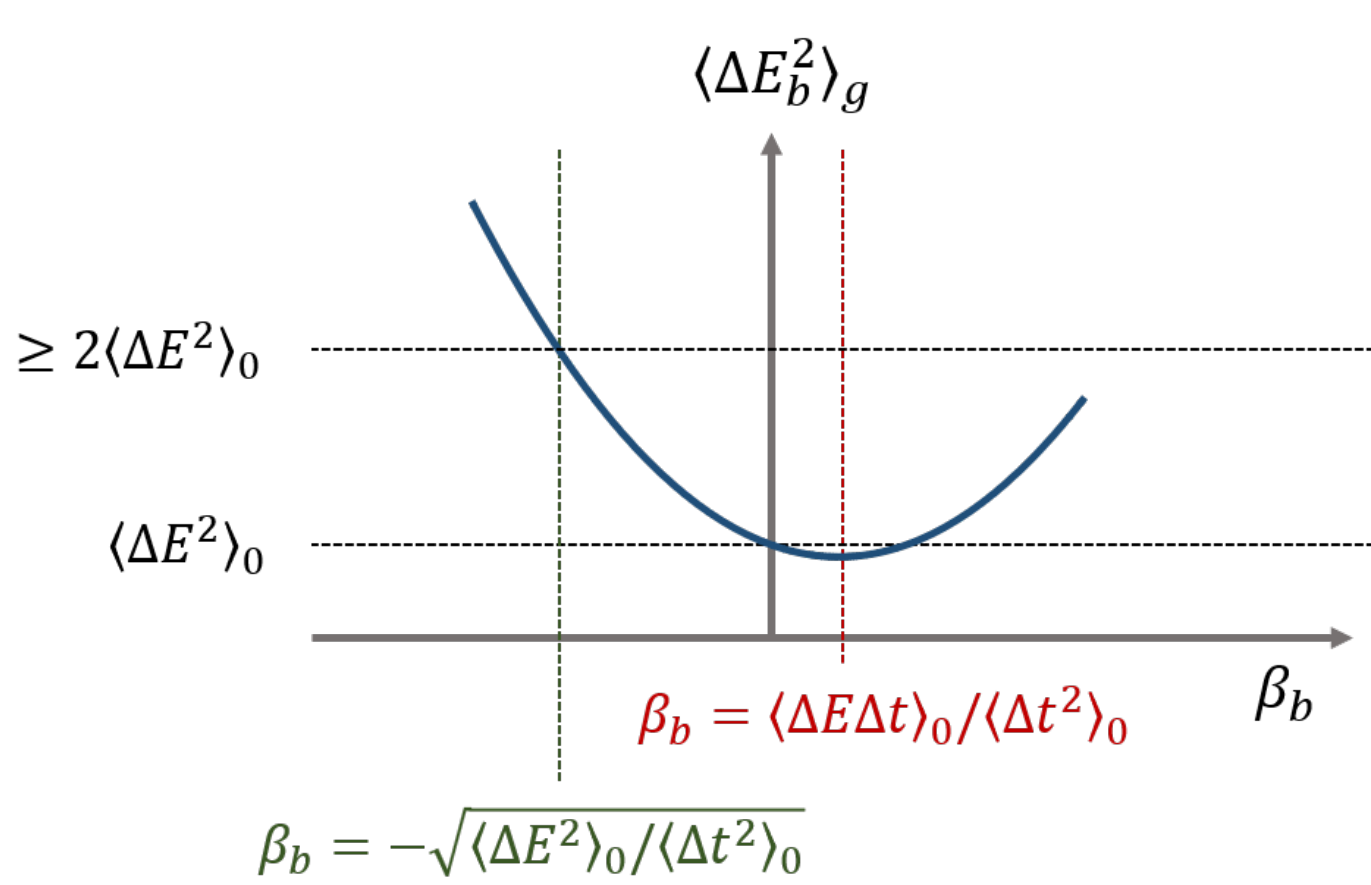

Figure S5. The barrier height variance $\langle \Delta E_b^2 \rangle_g$ (blue solid line) as a function of the rate $\beta_b$. The rates at which the variance $\langle \Delta E_b^2 \rangle_g$ reaches its minimum and exceeds twice the initial energy variance $\langle \Delta E^2 \rangle_0$ are indicated. The plot is drawn for positive covariance, but the case with negative covariance can be treated similarly.

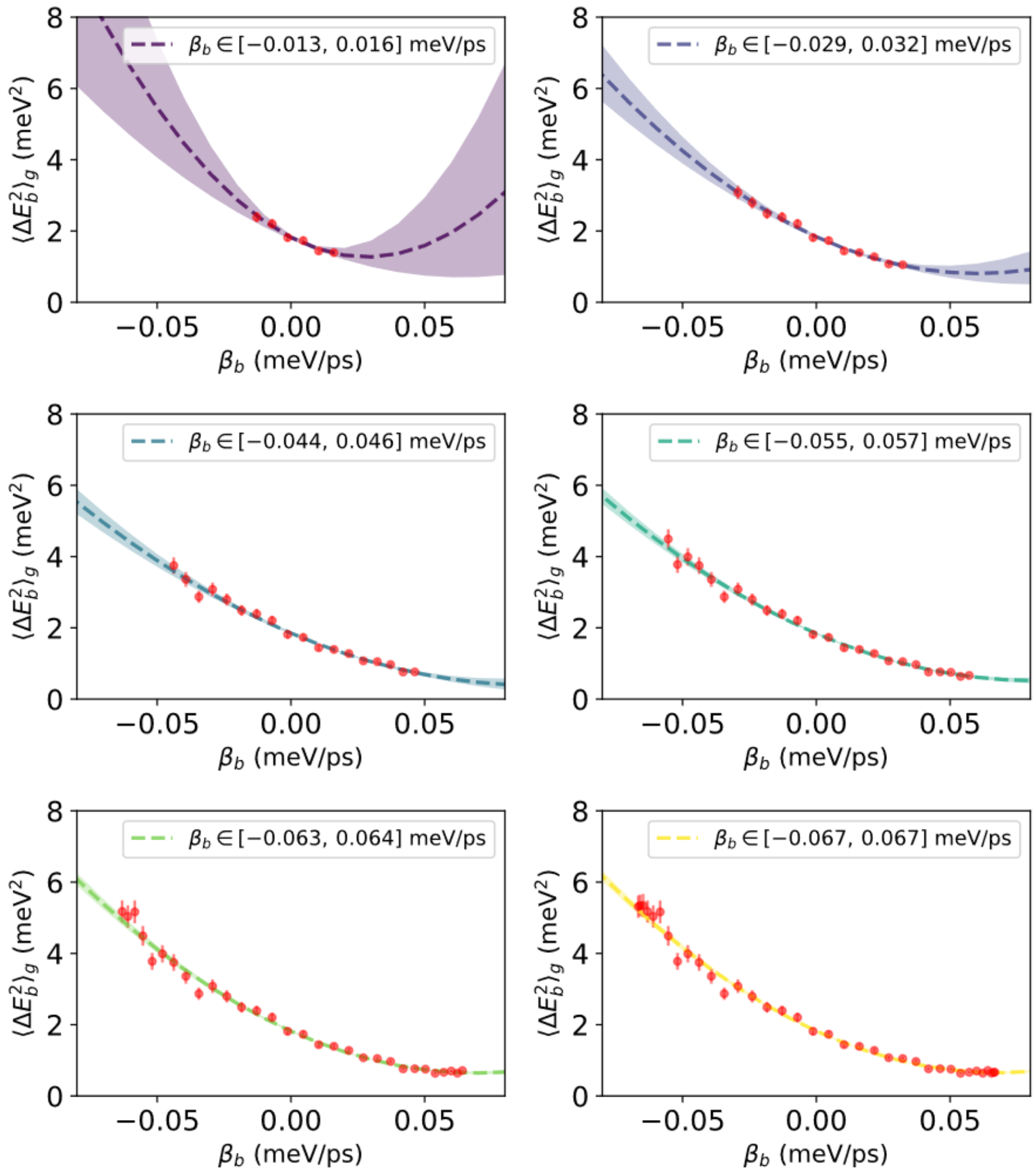

Figure S6. Experimental values of $\langle \Delta E_b^2 \rangle_g$ are shown as red circles, and the corresponding fits to Eq. (2) in the main text are shown as dashed line. The data are cropped from the full dataset of the red circles in Fig. 3(c), and each plot displays a difference range of $\beta_b$. The number of data points increases by 6 in each successive plot.

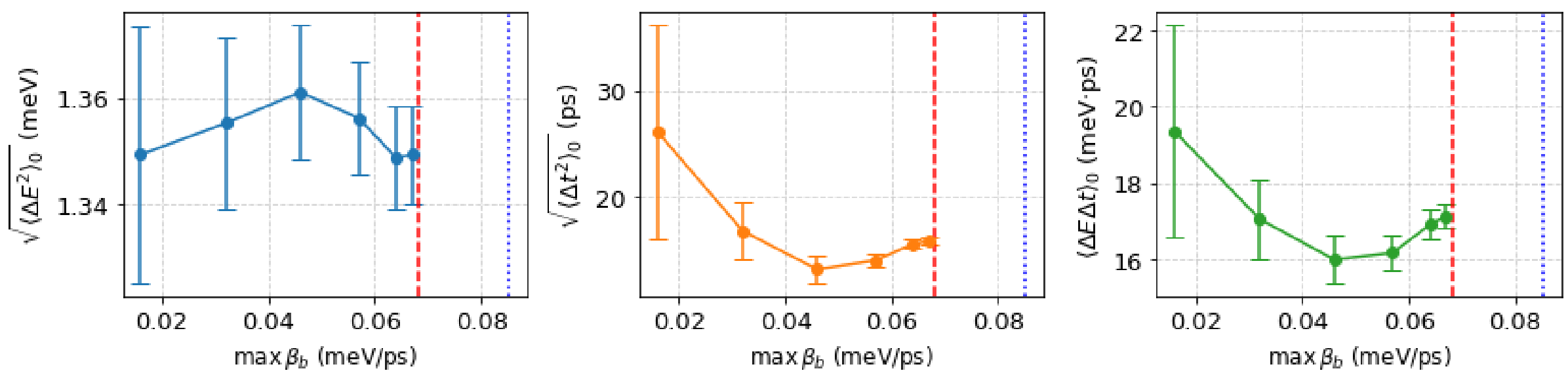

Figure S7. Extracted parameters $[\langle \Delta E^2 \rangle_0]^{1/2}$ (left panel) $[\langle \Delta t^2 \rangle_0]^{1/2}$ (middle panel), and $\langle \Delta E \Delta t \rangle_0$ (right panel), along with their errors from Fig. S6, plotted as functions of the maximum of the $\beta_b$ range. Red dashed lines show $\beta_b = \langle \Delta E \Delta t \rangle_0 / \langle \Delta t^2 \rangle_0$, and blue dotted lines show $\beta_b = [\langle \Delta E^2 \rangle_0 / \langle \Delta t^2 \rangle_0]^{1/2}$.

Given that the two transconductance regions [the upper red and lower blue regions in Fig. S8(a)] share the same range of ramping speed $\beta$, we compare the barrier height variance extracted from these regions, as shown in Fig. S8(b). The similarity of the extracted values (blue and red circles), which fit well to a single quadratic curve, demonstrates the robustness of our method. We note that, in the main text and the Supplementary Material, only the upper transconductance region is used to extract the barrier height variance.

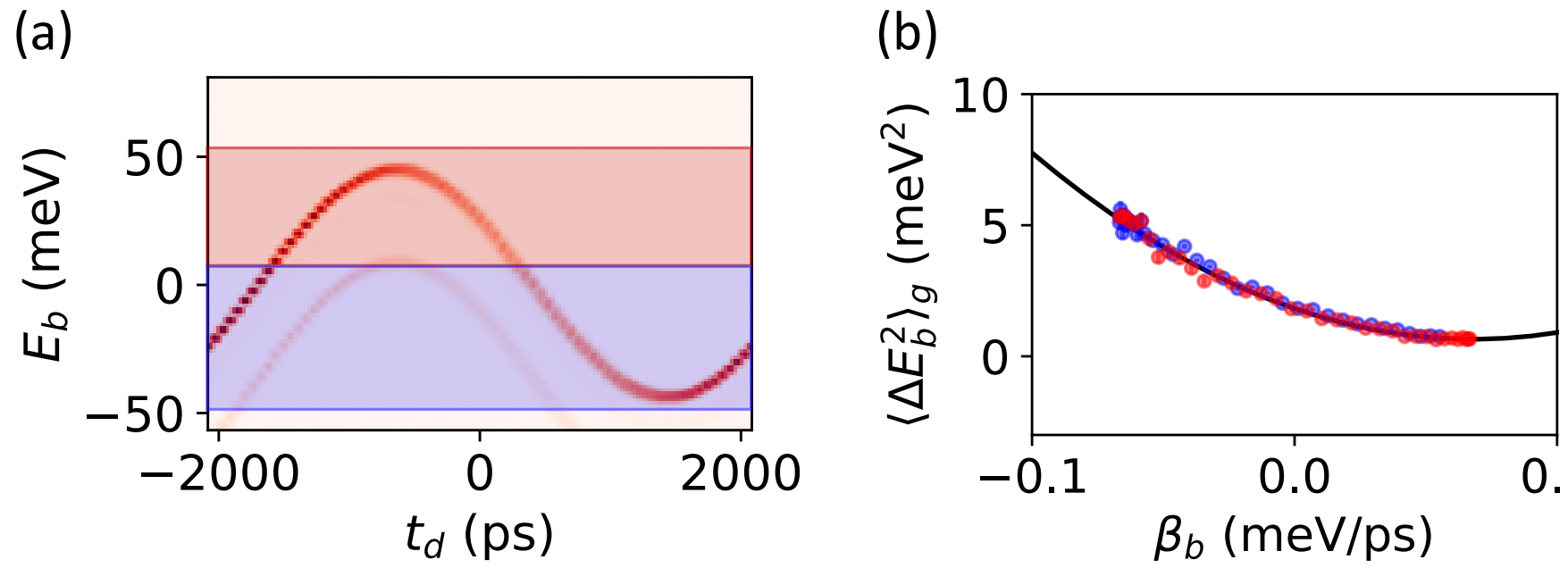

Figure S8. (a) Transconductance $g_b$ as a function of $t_d$ and $E_b$, measured using a sinusoidal waveform applied to the detector barrier [same as Fig. 3(a)]. (b) Barrier height variance $\langle \Delta E_b \rangle_g$ obtained from (a), shown in red and blue circles, extracted from the upper red and blue regions in (a), respectively.

## S8. Optimal time required for parameter extraction

In this section, we investigate the optimal time required to accurately extract parameters. For this purpose, we first analyze an optimal time-delay interval ($\delta t_d$) for extracting parameters. We next investigate an optimal barrier-height interval ($\delta E_b$) for extracting parameters and analyze errors arising from a finite $\delta E_b$. Here, the interval denotes the distance between the data points. From these analyses, we obtain the optimal number of data points to obtain the transconductance $g_b$, which is a function of the time delay $t_d$ and the barrier height $E_b$, and estimate the optimal time required for reliable parameter estimation.

We extract the barrier height variance $\langle \Delta E_b^2 \rangle_g$ for various time delay intervals $\delta t_d$ [Fig. S9(a)] at the P3 location in Fig. S4(b) of the pump device 2, measured with the detector frequency $f_b =$ 480 MHz. We note that the number of data points along $t_d$ can be obtained by dividing the period $1/f_b \simeq 2.1$ ns by $\delta t_d$. In this analysis, we keep the gate-voltage sweep interval fixed. The corresponding fit values [to Eq. (2) of the main text] are shown as a function of $\delta t_d$ in Fig. S9(b). We note that the first two red squares at $\delta t_d$ = 5.8 and 11.6 ps are extracted from measurements, which provide nearly unchanged parameters within error bars. We further increase $\delta t_d$ by cropping the experimental dataset used for the red square at $\delta t_d$ = 5.8 ps into several subsets (black squares), and investigate how the fit parameters vary as a function of the effective measurement time. The fit parameters remain nearly insensitive down to $\delta t_d \sim 57.9$ ps (37 data points), with small deviations down to $\delta t_d \sim 208.3$ ps (11 data points). From these results, we conclude that 37 data points along the time delay $t_d$ domain are sufficient to extract parameters.

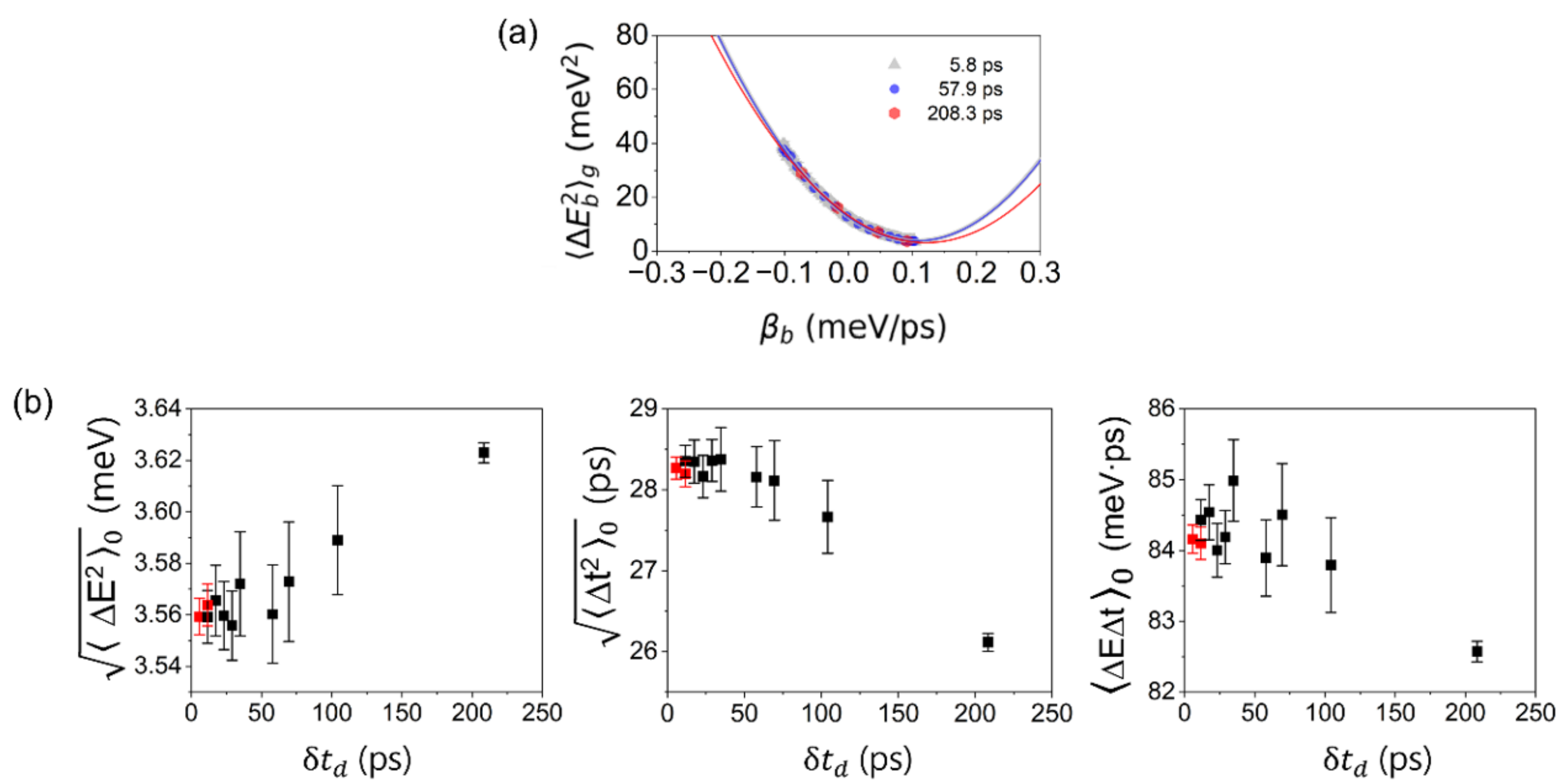

Figure S9. (a) Barrier height variance $\langle \Delta E_b^2 \rangle_g$ as a function of the barrier rate $\beta_b$ for various time delay intervals $\delta t_d$ = 5.8, 57.9, and 208.3 picoseconds. (b) Extracted parameters $[\langle \Delta E^2 \rangle_0]^{1/2}$ (left panel), $[\langle \Delta t^2 \rangle_0]^{1/2}$ (middle panel), and $\langle \Delta E \Delta t \rangle_0$ (right panel), along with their uncertainties, plotted as a function of $\delta t_d$.

We perform a similar analysis while varying the barrier height interval $\delta E_b$. This corresponds to varying the gate-voltage interval (spacing) in experiment. The data obtained at the P3 location in Fig. S4(b) of the pump device 2 is used for the analysis and measured with a barrier height interval $\delta E_b$ of 0.25 meV. By cropping this dataset, the data with larger intervals (0.5 meV, 0.75 meV, …) are created. From fits to the created datasets, we extract the energy variance, temporal variance, and energy-time covariance. Below, we estimate the optimal barrier height interval $\delta E_b$ by analyzing the extracted energy variance (width).

Figure S10 shows the measured energy width as a function of the barrier height interval $\delta E_b$. In the plot, both the measured energy width and the barrier height interval are normalized to the reference measured energy width (the energy width measured at the minimum barrier height interval, 3.56 meV). Therefore, from this plot, we can analyze how the error ($y$-axis) in the measured energy width changes as the ratio ($x$-axis) between the energy interval and the reference energy width varies. From the trend of this plot, we find that reliable parameters can be obtained when the ratio ($x$-value) is 0.1, corresponding to $\delta E_b$ = 0.356 meV (indicated by the blue vertical line in the plot). The error at this ratio is estimated to be ~1%. Here, we note that, in these data sets, the extracted temporal variance and energy–time covariance do not significantly change when $\delta E_b$ varies.

In addition to the analysis of the optimal barrier height interval, we quantitatively evaluate the errors arising from the finite energy interval in Figs. 3 and 4 of the main text. The corresponding ratios ($x$-values) are indicated by the red (the ratio used in Fig. 3) and blue (the ratio used in Fig. 4) vertical lines in Fig. S10. The errors are estimated as 4% and 1% for Figs. 3 and 4, respectively, and we have reflected these errors in the measured energy variance and computed area $A_0$.

Based on the above analyses, we approximately estimate the optimal number of data points required for accurate parameter extraction from the transconductance $g_b$. We note that $g_b$ is a function of the time delay $t_d$ and the barrier height $E_b$. Along the $E_b$ domain, from the earlier discussion, we estimate that the data points spaced by 0.1 times the barrier height width of the transconductance $g_b$ are optimal. To capture the whole finite region of $g_b$, a range extending over 15 times the width is considered optimal. From these analyses, we estimate that 150 points [= (width×15)/(width/0.1)] along the domain of $E_b$ are considered optimal. Combining this with the earlier finding that the number of time-delay points can be reduced to ~37 points, we estimate that 150 × 37 = 5550 data points are required for obtaining the transconductance $g_b$. The corresponding measurement time is 7.4 minutes, considering a measurement time of 0.08 seconds per single data point and that regions where the transconductance $g_b$ is zero are identified through a preliminary measurement using a large barrier height interval. This preliminary measurement takes additional time but will be less costly than, or at most comparable to, acquiring 5550 data points. Such measurement can also be performed using machine-learning techniques for further optimization [5].

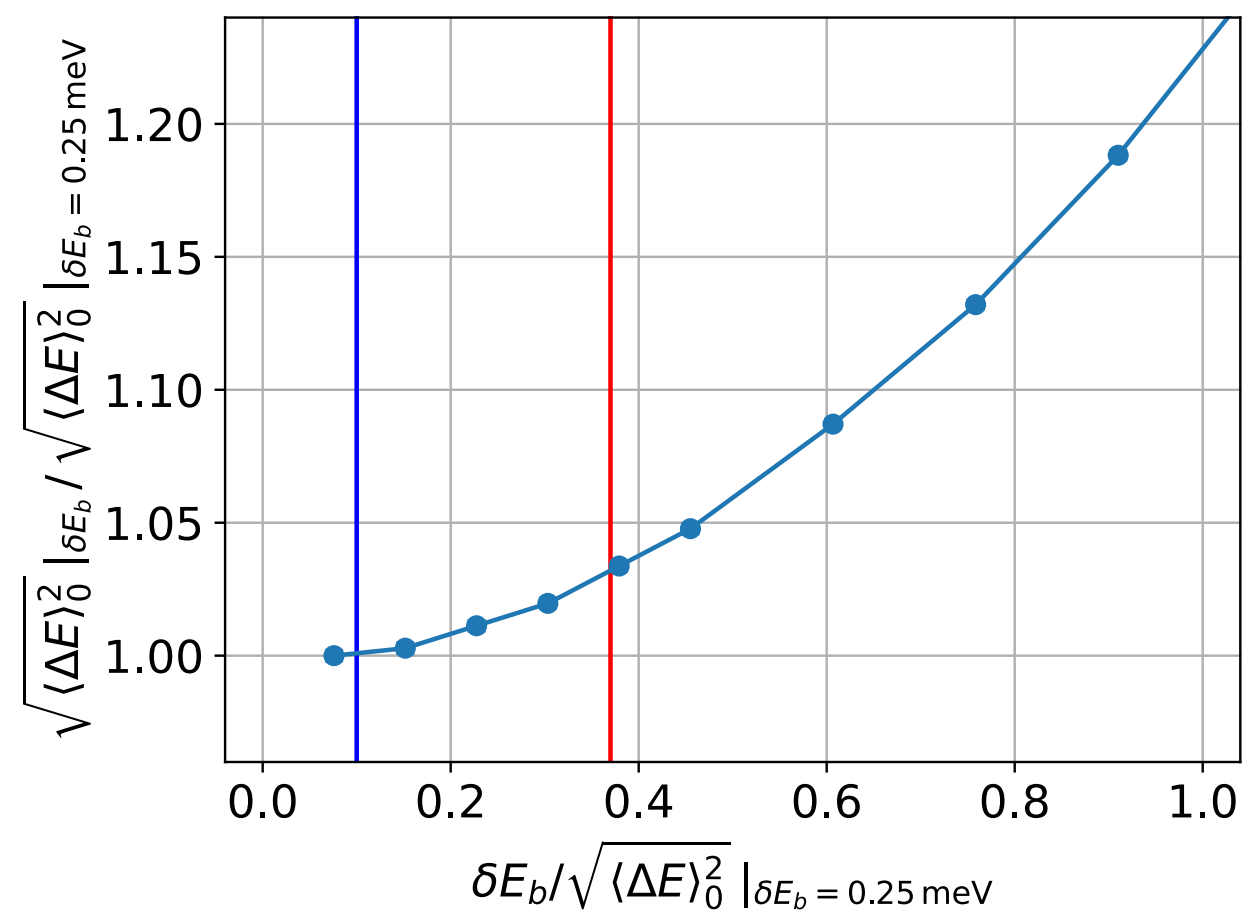

Figure S10. Extracted energy widths $\langle \Delta E^2 \rangle_0$ as a function of the barrier height interval (blue circles). Both *x*- and *y*-axes are normalized to the reference energy (the energy width measured at $\delta E_b = 0.25$ meV). The blue and red vertical lines denote the normalized barrier height intervals used in Figs. 3 and 4 of the main text, respectively.

## S9. The effective phase-space area computed from the covariance matrix

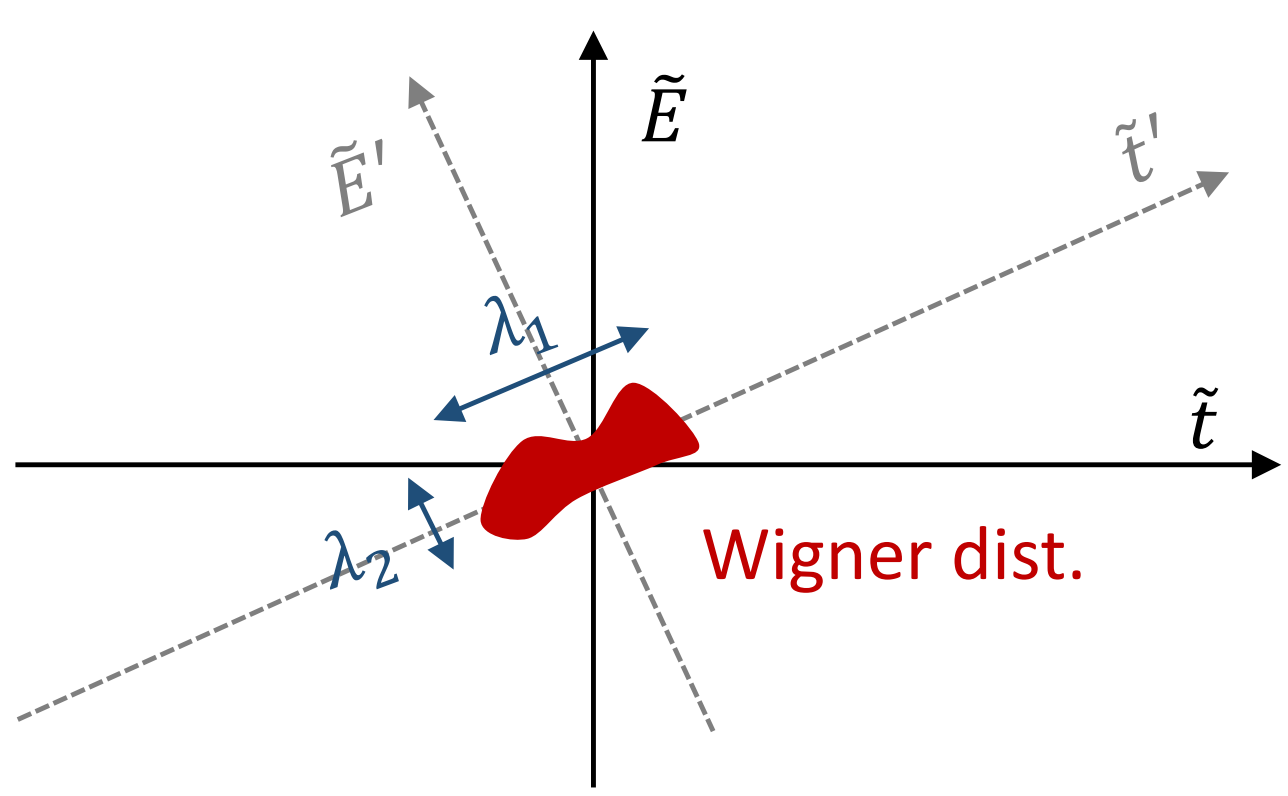

Figure S10. A Wigner distribution (red region), with standard deviations $\lambda_i$ (square roots of the eigenvalues of $\tilde{\Sigma}_0$) along the principal axes $\tilde{t}'$ and $\tilde{E}'$, which are determined by the eigenvectors $\vec{u}_i$ of $\tilde{\Sigma}_0$. They are shown in dimensionless coordinates.

Here, we compute the effective phase-space area and tilted angle from the covariance matrix. First, to make quantities dimensionless, we introduce $\tilde{E} = E/\Lambda_E$, $\tilde{t} = t/\Lambda_t$, (see the solid black axes in Fig. S11) and $\widetilde{W} = \Lambda_E \Lambda_t W$.

In the $(\tilde{E},\ \tilde{t})$ basis, the covariance matrix is

$$\tilde{\Sigma}_0 = \langle \begin{pmatrix} \Delta\tilde{E} \\ \Delta\tilde{t} \end{pmatrix} (\Delta\tilde{E}\ \ \Delta\tilde{t}) \rangle_0 = \begin{pmatrix} \langle \Delta\tilde{E}^2 \rangle_0 & \langle \Delta\tilde{E}\Delta\tilde{t} \rangle_0 \\ \langle \Delta\tilde{E}\Delta\tilde{t} \rangle_0 & \langle \Delta\tilde{t}^2 \rangle_0 \end{pmatrix}. \tag{S9}$$

We introduce principal axis coordinates $(\tilde{E}', \tilde{t}')$ via the rotation (see the dashed gray coordinates in Fig. S11)

$$\begin{pmatrix} \tilde{E}' \\ \tilde{t}' \end{pmatrix} = R^T \begin{pmatrix} \tilde{E} \\ \tilde{t} \end{pmatrix}, \tag{S10}$$

where $R = [\vec{u}_1, \vec{u}_2]$ is the orthonormal matrix whose columns $\vec{u}_1$ and $\vec{u}_2$ are eigenvectors of $\tilde{\Sigma}_0$. The covariance matrix $\tilde{\Sigma}_0'$ in these principal-axis coordinates becomes

$$\tilde{\Sigma}_0' = \left\langle R^T \begin{pmatrix} \Delta\tilde{E} \\ \Delta\tilde{t} \end{pmatrix} (\Delta\tilde{E}\ \ \Delta\tilde{t}) R \right\rangle = R^T \tilde{\Sigma}_0 R. \tag{S11}$$

The covariance matrix is diagonalized such that $\tilde{\Sigma}_0' = \mathrm{diag}(\lambda_1^2, \lambda_2^2)$, where $\lambda_1$ and $\lambda_2$ are standard deviations (widths) of the distribution along the principal axis. Hence, the off-diagonal term of the covariance vanishes. We define the effective phase-space area as the product of the uncertainties of electrons along the principal axes,

$$\tilde{A}_0 = \lambda_1 \lambda_2 = \sqrt{\det \tilde{\Sigma}_0}, \tag{S12}$$

where we use that the determinant of a $2 \times 2$ matrix equals the product of its eigenvalues. This area is equivalent to that of an ellipse with width $2\lambda_1/\sqrt{\pi}$ and height $2\lambda_2/\sqrt{\pi}$. The dimensionful effective phase-space area becomes

$$A_0 = \Lambda_E \Lambda_t \tilde{A}_0 = \sqrt{\det \Sigma_0}, \tag{S13}$$

where $\Sigma_0$ is Eq. (3) of the main text. Equation (S13) is the same as Eq. (4) of the main text. The area is represented in ellipse in Figs. 3(e), 4(b), and 4(c) of the main text. Here, Eq. (S13) is invariant under choice of $\Lambda_E$ and $\Lambda_t$.

The dimensionful slope of the principal axes can be computed from the eigenvectors $\vec{u}_1$ and $\vec{u}_2$ and the factors $\Lambda_E$ and $\Lambda_t$. Here, this slope depends on choice of $\Lambda_E$ and $\Lambda_t$. The interpretation of Eq. (S13) as a phase-space area becomes particularly intuitive when $W_0$ has an elliptical shape corresponding to a bivariate Gaussian distribution (see Sec. S10).

From the Robertson-Schrödinger uncertainty principle, this effective area is bounded by [6,7],

$$\det \Sigma_0 = \langle \Delta E^2 \rangle_0 \langle \Delta t^2 \rangle_0 - \langle \Delta E \Delta t \rangle_0^2 \geq (\hbar/2)^2, \tag{S14}$$

where $\hbar = h/(2\pi)$ is the reduced Planck constant and $h$ is the Planck constant. Then, $A_0$ of Eq. (S13) [Eq. (4) of the main text] follows

$$A_0 \geq \hbar/2. \tag{S15}$$

## S10. Elliptical distribution

In this section, we introduce the elliptical form of a bivariate Gaussian distribution, used for plotting the Wigner distribution in Figs. 3(e), 4(b), 4(c), and S14. We compute the purity of this distribution and show that the effective area $A_0$ of Eq. (S13) [Eq. (4) of the main text] reaches the quantum limit $\hbar/2$ when the purity equals 1. We then represent the Wigner distribution as a density matrix and relate the statistical weights (i.e., the probabilities of pure states in the mixture) to the effective area $A_0$ within the elliptical approximation.

The bivariate Gaussian Wigner distribution can be expressed in terms of the covariance matrix $\Sigma_0$ as

$$W_0(E,t) = \frac{1}{2\pi\sqrt{\det \Sigma_0}} \exp\left[-\frac{1}{2}\begin{bmatrix} E \\ t \end{bmatrix}^T \Sigma_0^{-1} \begin{bmatrix} E \\ t \end{bmatrix}\right] \tag{S16}$$

$$= \frac{1}{2\pi\sigma_{E0}\sigma_{t0}\sqrt{1-r_0^2}} \exp\left[-\frac{1}{2(1-r_0^2)}\left(\frac{E^2}{\sigma_{E0}^2} + \frac{t^2}{\sigma_{t0}^2} - \frac{2r_0 E t}{\sigma_{E0}\sigma_{t0}}\right)\right]. \tag{S17}$$

In the second line, the distribution is expressed using the energy width $\sigma_{E0}$ $(= \sqrt{\langle \Delta E^2 \rangle_0})$, the time width $\sigma_{t0}$ $(= \sqrt{\langle \Delta t^2 \rangle_0})$, and the correlation coefficient $r_0$ $(= \langle \Delta E \Delta t \rangle_0 / \sqrt{\langle \Delta E^2 \rangle_0 \langle \Delta t^2 \rangle_0})$.

The purity $\gamma_0$ is defined as the trace of the square of the density matrix and is used as a measure on quantum states giving information on how much the state is mixed. This can be computed from the Wigner distribution as $W_0$ [8]

$$\gamma_0 = 2\pi\hbar \int_{-\infty}^{\infty} dE \int_{-\infty}^{\infty} dt\, W_0^2. \tag{S18}$$

Under the elliptical approximation of Eq. (S16), the purity becomes

$$\gamma_0 = \frac{\hbar}{2}[\det \Sigma_0]^{-1/2} = \frac{\hbar}{2}\frac{1}{A_0}. \tag{S19}$$

This is bounded by the Robertson-Schrödinger uncertainty relation given in Eq. (S14), ensuring $\gamma_0 \le 1$. For the pure state of $\gamma_0 = 1$, the area $A_0$ equals $\hbar/2$, irrespective of the value of the covariance $\langle \Delta E \Delta t \rangle_0$.

We now represent the Wigner distribution as the density matrix

$$\rho_0 = \sum_{i=1}^{\infty} p_i |\psi_i\rangle\langle\psi_i|, \tag{S20}$$

within the elliptical approximation, where $|\psi_i\rangle$ are normalized pure states with statistical weight $p_i$. The form of Eq. (S20) can be obtained using (1) the Wigner–Weyl transform relation between the Wigner distribution $W_0$ of Eq. (S16) and the density matrix $\rho_0$ and (2) the Mehler Kernel [9]. The resulting weight factor is

$$p_i = \left(\frac{2\hbar}{2A_0 + \hbar}\right)\left(\frac{2A_0 - \hbar}{2A_0 + \hbar}\right)^{i-1}, \tag{S21}$$

which depends on the effective area $A_0$ defined in Eq. (S13). It follows that $\sum_{i=1}^{\infty} p_i = 1$ and that the purity is given by $\gamma_0 = \sum_{i=1}^{\infty} p_i^2$ [see Eqs. (S19) and (S21)]. A pure state ($p_1 = 1$) is obtained when $A_0 = \hbar/2$, as expected. The largest weight $p_1$ is

$$p_1 = \frac{2\hbar}{2A_0 + \hbar} \simeq \frac{\hbar}{A_0} \quad \text{for } A_0 \gg \hbar/2, \tag{S22}$$

which coincides with the peak value of $W_0$ [see Eq. (S16)] and agrees with Ref. [3]. Equation (S22) shows that the largest weight $p_1$ can be estimated directly from $A_0$ of Eq. (S13).

**S11. Tomographic reconstruction**

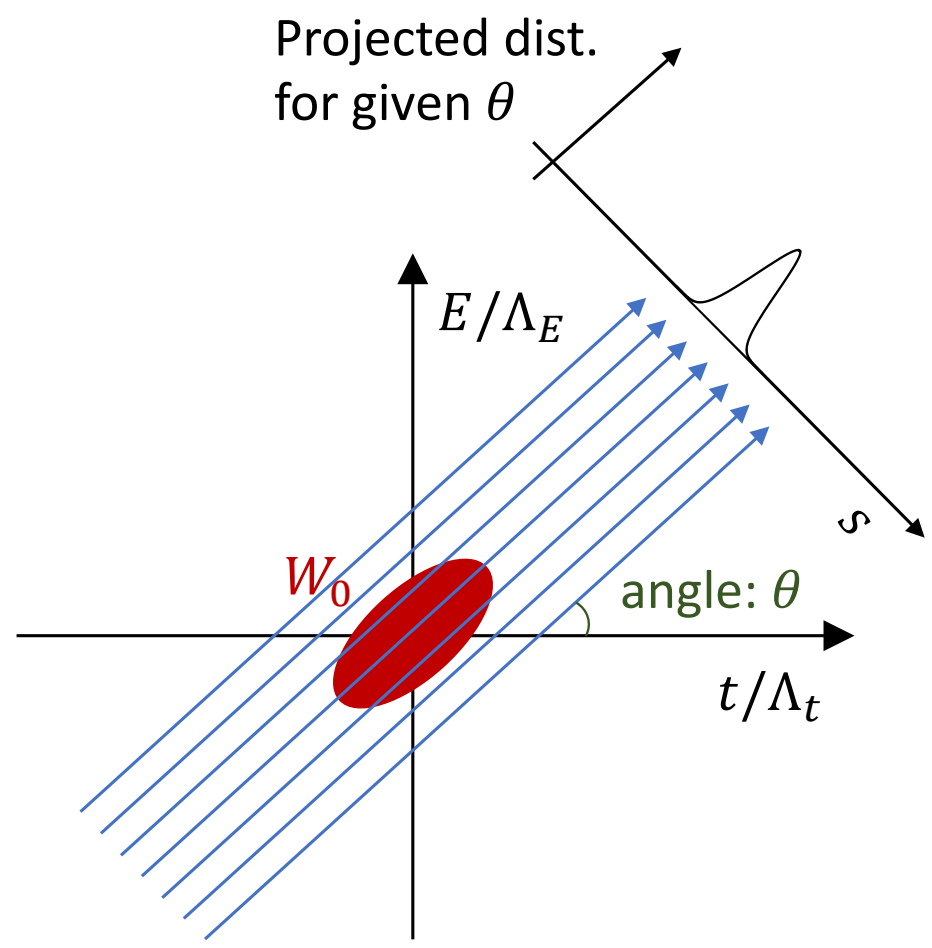

Figure S11. The Wigner distribution $W_0$ (red oval) projected along the line (blue arrows) at an angle $\theta$. The projected distribution for given $\theta$ is shown as the black curve in the upper right panel. $\Lambda_t$ and $\Lambda_E$ are normalized values used to make quantities dimensionless.

We introduce tomographic methods to obtain reconstructed Wigner distributions in Figs. 3 and 4 of the main text. We then compare the results with the bivariate Gaussian distribution with the parameters obtained from $\Sigma_0$ and investigate an error by a finite tomographic angle.

We employ the tomographic method to reconstruct the Wigner distribution as in Fletcher et al. [3] and Kim et al. [10]. We first outline the transformation method as schematically shown in Fig. S12. Our target to reconstruct is a Wigner distribution $W_0$ (denoted in time-energy phase space, shown as the red oval). We achieve this by projecting the Wigner distribution (see below for the experimental method) with an angle $\theta$ along the $s$ axis (upper right panel), which is perpendicular to the projection lines (blue arrows). By collecting the projected distributions while varying $\theta$, we obtain a sinogram, which is the Radon-transformed data of the target. Using the inverse Radon transformation, we finally obtain the target of the Wigner distribution.

We review the detailed experimental method to obtain the projected distribution and the sinogram [3,10]. The projected distribution can be obtained by measuring a transmitted current $I_T$ through the detector barrier at a rate $\beta_b$. For a linearly time-varying potential (a hyperbolic tangent shape is typically used as implemented in the experiment and described in the main text), the transmission probability $P_T$ (the normalized value of $I_T$ with pumping current $ef$) follows [11]

$$P_T = \int_{-\infty}^{\infty} dt \int_{-\infty}^{\infty} dE \, W_0(E, t) \, T(E - E_b - \beta_b(t - t_d)), \tag{S23}$$

where $T$ represents the transmission probability of a monochromatic wave through the barrier. $E_b$ and $t_d$ denote the DC barrier height and time delay of the detector barrier, respectively. When the energy width of $dT/dE$ is negligible compared to the energy scale of $W_0$, $T$ can be approximated as a step function. In this case, the derivative of $P_T$ provides the projected distribution along a given derivative direction. The finite width of $dT/dE$ gives an error in measurement [11]. To make phase-space coordinate dimensionless, we introduce the normalized parameters, $\Lambda_E$ ($E \to E/\Lambda_E$ and $E_b \to E_b/\Lambda_E$) and $\Lambda_t$ ($t \to t/\Lambda_t$). In this coordinate, the projection angle $\theta$ is related to the rate $\beta_b$ by $\theta = \tan^{-1}(\beta_b/\beta_0)$, where $\beta_0 = \Lambda_E/\Lambda_t$. $\beta_0$ is empirically chosen in the experiment. As in Ref. [3], we use a hyperbolic tangent waveform to the detector,

$$V_b(t) = V_{\mathrm{AC}} \tanh[A_0 \tan\theta \sin[2\pi f_b(t - t_d)]] + V_{\mathrm{DC}}. \tag{S24}$$

In this waveform, $\beta_0$ is determined by $A_0$, $V_{\mathrm{AC}}$, and gate conversion factor.

By collecting the projected distribution ($\simeq dP_T/ds$) ($s$ is the coordinate perpendicular to the projection lines, see Fig. S12) while varying the projection angle $\theta$, barrier height $E_b$, and time delay $t_d$, we obtain the sinogram (Fig. S13). Due to experimental limitations, the projection angles above 85° or below −85° cannot be reached. This limitation may arise from the finite bandwidth of AWG or signal damping at large $\beta_b$. Using the inverse Radon transformation (implemented via the iradon function in the skimage python library) and reverting the dimensionless coordinate back to their dimensional form, we reconstruct the Wigner distribution (Fig. S14).

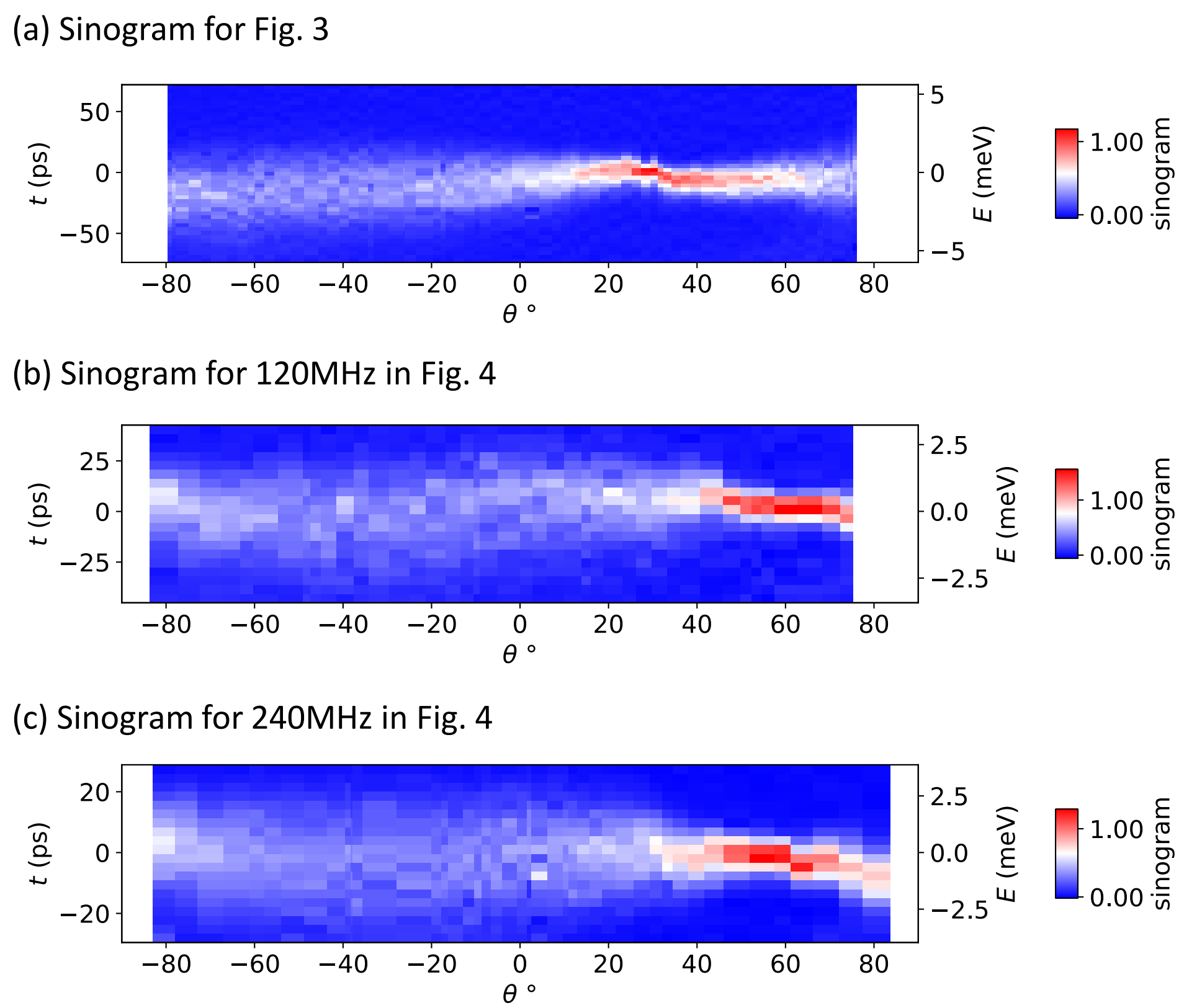

Figure S12. Sinograms as a function of the projection angle $\theta$ and the dimensionless coordinate $s$. The corresponding time axis (for $\theta = 90°$, where the sinogram represents the arrival time distribution) and energy axis (for $\theta = 0°$, where the sinogram represents the energy distribution) are shown in the plot. The sinograms are obtained for the generated electrons under the following conditions: Fig. 3, Fig. 4(b), and Fig. 4(c) of the main text, corresponding to panels (a), (b), and (c), respectively. In (a) and (c), $\beta_0 = 0.09$ meV/ps is used, while in (b), $\beta_0 = 0.05$ meV/ps is used.

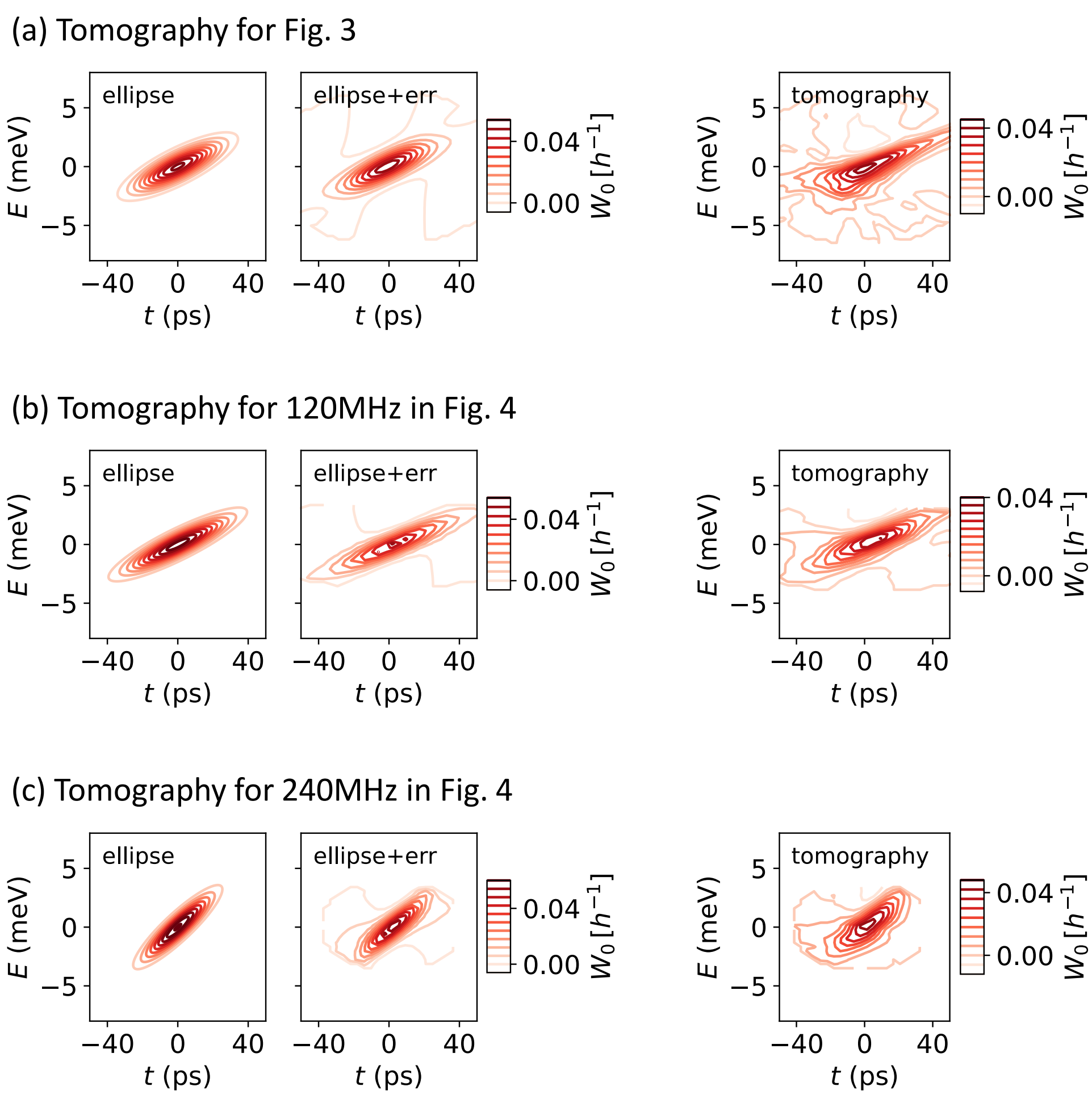

Figure S13. Wigner distribution $W_0$ of generated electrons in the time-energy phase space. The right panels show the results obtained through the tomographic procedure (see sinogram in Fig. S13). The left panels present the bivariate Wigner distributions with the parameters obtained in the main text. The middle panels account for errors due to a finite tomographic angle in the right panel (see the text for detail). The results correspond to the generated electrons under the conditions of (a) Fig. 3 and (b, c) Fig. 4 of the main text. For (b) and (c), the pumping frequency are 120 MHz and 240 MHz, respectively.

To independently evaluate errors caused by the limited range of tomographic angle, we further compute the sinogram of the bivariate Gaussian Wigner distributions using the same tomographic angles and the same range of $s$ coordinate as in the experimental tomographic reconstruction. The inverse Radon-transformed results, accounting for these errors, are shown in the right sides of the left panels of Fig. S14.

We found overall agreement between the tomographic reconstruction results and the results obtained from the bivariate Gaussian Wigner distribution using the extracted parameters in the main text. The results support our time-resolved measurements utilizing a barrier height variance of the transconductance. Reconstruction errors arise from the limited range of tomographic angles and the presence of noises, which we estimate, particularly higher at higher values of $\beta_b$.

**Supplementary references**